%% file: main.tex
\documentstyle[12pt,epsf]{article}
\begin{document}
\def\susy23{S{\small USY23}}
\def\D#1{{\bf D#1}}
\def\B#1{{\bf B#1}}
\def\C#1{{\bf C#1}}
\def\slepton{\widetilde \ell}
\def\sl{{\widetilde \ell}^{-}}
\def\slb{{\widetilde \ell}^{+}}
\def\slr{\slepton_{R}}
\def\sll{\slepton_{L}}
\def\squark{\widetilde q}
\def\msl{m_{\slepton}}
\def\msq{m_{\squark}}
\def\msll{m_{\sll}}
\def\mslr{m_{\slr}}
\def\photino{\widetilde \gamma}
\def\zino{{\widetilde{Z}}}
\def\wino{{\widetilde{W}}}
\def\sfermion{\widetilde f}
\def\sf{\widetilde f}
\def\gluino{\widetilde g}
\def\sfb{\widetilde{\overline f}}
\def\sneu{\widetilde \nu}
\def\snel{\widetilde \nu_{e}}
\def\snmu{\widetilde \nu_{\mu}}
\def\sntau{\widetilde \nu_{\tau}}
\def\sele{\widetilde e}
\def\smuon{\widetilde \mu}
\def\sqk{\widetilde q}
\def\ser{\sele_{R}}
\def\sel{\sele_{L}}
\def\smr{\smuon_{R}}
\def\sml{\smuon_{L}}
\def\sql{\squark_{L}}
\def\sqr{\squark_{R}}
\def\msql{m_{\sql}}
\def\msqr{m_{\sqr}}
\def\msf{m_{\sfermion}}
\def\msn{m_{\sneutrino}}
\def\msne{m_{\snel}}
\def\msnm{m_{\snmu}}
\def\msnt{m_{\sntau}}
\def\msg{m_{\gluino}}
\def\smu{\widetilde \mu}
\def\stau{\widetilde \tau}
\def\staul{\stau_{L}}
\def\staur{\stau_{R}}
\def\su{\widetilde{u}}
\def\sd{\widetilde{d}}
\def\sc{\widetilde{c}}
\def\ss{\widetilde{s}}
\def\sul{\su_{L}}
\def\sdl{\sd_{L}}
\def\sur{\su_{R}}
\def\sdr{\sd_{R}}
\def\scl{\sc_{L}}
\def\ssl{\ss_{L}}
\def\scr{\sc_{R}}
\def\ssr{\ss_{R}}
\def\sclr{\sc_{L,R}}
\def\sslr{\ss_{L,R}}
\def\sulr{\su_{L,R}}
\def\sdlr{\sd_{L,R}}
\def\msul{m_{\sul}}
\def\msur{m_{\sur}}
\def\msdl{m_{\sdl}}
\def\msdr{m_{\sdr}}
\def\mscl{m_{\scl}}
\def\mscr{m_{\scr}}
\def\mssquarkl{m_{\ssl}}
\def\mssr{m_{\ssr}}
\def\msclr{m_{\sclr}}
\def\msulr{m_{\sulr}}
\def\msdlr{m_{\sdlr}}
\def\msslr{m_{\sslr}}
\def\mch{m_{H^{+}}}
\def\mA{m_{A}}
\def\mH{m_{H}}
\def\st{\widetilde{t}}
\def\sb{\widetilde{b}}
\def\mstaul{m_{\stau_{1}}}
\def\mstauh{m_{\stau_{2}}}
\def\ddf{{\rm d}}
\def\bino{\widetilde{B}}
\def\higgsino{\widetilde{H}^{0}}
\def\sz1{{\widetilde{Z}}_{1}}
\def\szs{{\widetilde{Z}}_{2}}
\def\szt{{\widetilde{Z}}_{3}}
\def\szf{{\widetilde{Z}}_{4}}
\def\szk{{\widetilde{Z}}_{k}}
\def\swl{{\widetilde{W}}_{1}}
\def\swh{{\widetilde{W}}_{2}}
\def\swi{{\widetilde{W}}_{i}}
\def\mszk{m_{\szk}}
\def\mswi{m_{\swi}}
\def\mse{m_{\sele}}
\def\mser{m_{\ser}}
\def\msel{m_{\sel}}
\def\msmr{m_{\smr}}
\def\msml{m_{\sml}}
\def\msz1{m_{\sz1}}
\def\mszs{m_{\szs}}
\def\mszt{m_{\szt}}
\def\mszf{m_{\szf}}
\def\mswl{m_{\swl}}
\def\mswh{m_{\swh}}
\def\mbino{m_{\bino}}
\def\gev{{\rm GeV}}
\def\tev{{\rm TeV}}
\def\rs{{\sqrt{s}}}
\def\tanbe{\tan\beta}
\def\nle{{\stackrel{<}{\sim}}}
\def\nge{{\stackrel{>}{\sim}}}
\def\goto{\rightarrow}
\def\bhd{{\hat{\beta'}}}
\def\half{{\frac{1}{2}}}
\def\mt{m_{t}}
\def\mb{m_{b}}
\def\mh{m_{h}}
\def\stl{\st_{1}}
\def\stls{\st^{*}_{1}}
\def\sth{\st_{2}}
\def\sbl{\sb_{1}}
\def\sbr{\sb_{R}}
\def\sbh{\sb_{2}}
\def\mst{m_{\st}}
\def\mstl{m_{\stl}}
\def\msth{m_{\sth}}
\def\msb{m_{\sb}}
\def\msbl{m_{\sbl}}
\def\msbh{m_{\sbh}}
\def\mz{m_{Z}}
\def\mw{m_{W}}
\def\tht{\theta_{t}}
\def\thb{\theta_{b}}
\def\tha{\theta_{\tau}}
\def\tew{\theta_{W}}
\def\sw{\sin^{2}\theta_{W}}
\def\cw{\cos^{2}\theta_{W}}
\def\muo{\mu_{\infty}}
\def\mgo{M_{\infty}}
\def\afo{A_{\infty}}
\def\mfo{m_{\infty}}
\def\mgut{M_{X}}
\def\mtil{\widetilde{m}}
\def\Xtil{\widetilde{X}}
\def\misEt{\slash\hspace{-8pt}E_{T}}
\def\misP{\slash\hspace{-8pt}P_{T}}
\def\rb{\slash\hspace{-6pt}R}
\def\pos{\slash\hspace{-6pt}p_{1}}
\def\pts{\slash\hspace{-6pt}p_{2}}
\def\prs{\slash\hspace{-6pt}p_{3}}
\def\mot{m_{12}^{2}}
\def\mor{m_{13}^{2}}
\def\mtr{m_{23}^{2}}
\def\mssql{\mtil^{2}_{Q_{1,2}}}
\def\mssqh{\mtil^{2}_{Q_{3}}}
\def\mssul{\mtil^{2}_{U_{1,2}}}
\def\mssuh{\mtil^{2}_{U_{3}}}
\def\mssdl{\mtil^{2}_{D_{1,2}}}
\def\mssdh{\mtil^{2}_{D_{3}}}
\def\mssl{\mtil^{2}_{L}}
\def\msse{\mtil^{2}_{E}}
\def\msh1{\mtil^{2}_{H_{1}}}
\def\msh2{\mtil^{2}_{H_{2}}}
\def\Ktil{\widetilde{K}}
\def\Itil{\widetilde{I}}
\def\lam{\lambda'_{131}}
\def\i{{\rm i}}
%
\input titleabs.tex

\vskip 0.8cm
%
\input cpchead.tex
\vskip 0.8cm
\topmargin -1cm
{\small
\input chap12.tex

\vskip 0.8cm
%
\input chap3.tex
\vskip 0.8cm
%
\input chap4.tex

%
\vskip 0.8cm
%
\input howto.tex
%
\vskip 0.8cm
%
\input summary.tex
%
%
\vskip 0.8cm
\input ack.tex
\vskip 0.8cm
%
\input ref.tex

%
\newpage
\vskip 0.8cm
\vskip 0.8cm
%
\input appa.tex

%
\vskip 0.6cm
%
\input appb.tex
\vskip 0.8cm
%
\input appc.tex

%
\vskip 0.8cm
%
\input appd.tex

\vskip 0.8cm
\input testrun.tex
}
\end{document}

%% file: titleabs.tex
\par
\vspace{10mm}
\begin{flushright}
KEK Preprint 97-211 \\
ITP-SU-97/04  \\
TMCP-97-1 \\
LAPP-EXP-97.07 \\
KEK CP-058 \\
hep-ph/9711283 \\
Oct, 1997
\end{flushright}
\vskip 3cm
\centerline{\large {\Large {\susy23}} v2.0:}
\centerline{\large an Event Generator for Supersymmetric Processes}
\centerline{\large at $e^+e^-$ Colliders }
\par
\vskip 0.3cm
{\small
\centerline{J. Fujimoto, K. Hikasa$^{a)}$, T. Ishikawa, M. Jimbo$^{b)}$, T. Kaneko$^{c)}$,}
\centerline{K. Kato$^{d)}$, S. Kawabata, T. Kon$^{e)}$, M. Kuroda$^{c)}$, Y. Kurihara, T. Munehisa$^{f)}$,}
\centerline{D. Perret-Gallix$^{g)}$, Y. Shimizu and H. Tanaka$^{h)}$} 
}
\vskip 0.2cm
{\footnotesize\it
\centerline{KEK, Oho, Tsukuba, Ibaraki 305, Japan}
\centerline{a) Tohoku University, Aoba-ku, Sendai 980-77, Japan}  
\centerline{b) Tokyo Management College, Ichikawa, Chiba 272, Japan}
\centerline{c) Meiji-Gakuin University, Kamikurata, Totsuka, Yokohama 244, Japan}
\centerline{d) Kogakuin University, Nishi-Shinjuku, Tokyo 160, Japan}
\centerline{e) Seikei University, Musashino, Tokyo 180, Japan}
\centerline{f) Yamanashi University, Takeda, Kofu 400, Japan}
\centerline{g) LAPP, B.P. 110, Annecy-le-Vieux F-74941 CEDEX, France}
\centerline{h) Rikkyo University, Nishi-Ikebukuro, Tokyo 171, Japan}
}
\par
\vskip 0.3cm
%
\vskip 0.2cm
\noindent
\unitlength 1mm
\begin{picture}(160,1)
\put(0,0){\line(1,0){160}}
\end{picture}
\par
%
\leftline{\small\bf Abstract}
{\footnotesize
{\large {\susy23}} is a Monte-Carlo package
for generating supersymmetric (SUSY) processes 
at $e^+e^-$ colliders. 
Twenty-three types of SUSY processes with 2 or 3 final state
particles 
at tree level are included in version 2.0.
{\susy23} addresses event simulation requirements at $e^+e^-$
colliders such as LEP.
Matrix elements are generated by {\tt GRACE} with the 
helicity amplitude method
for processes involving massive fermions.
The phase space integration of the matrix element 
gives the total and differential cross sections, 
then unweighted events are generated. 
Sparticle widths and decay branching ratios are calculated.
Each final state particle may then decay according to these
probabilities. Spin correlations are taken into account in the
decays of sparticles.
Corrections of initial state 
radiation (ISR) are implemented in two ways, one is based on the
electron structure function 
formalism and the second uses the parton shower algorithm called {\tt QEDPS}. 
Parton shower and hadronization of 
the final quarks are performed through an interface to 
{\tt JETSET}. 
}
\par\noindent
\begin{picture}(160,1)
\put(0,0){\line(1,0){160}}
\end{picture}
\par

%% file: cpchead.tex
%
\noindent
\begin{minipage}{7.6cm}
\leftline{\small\bf PROGRAM SUMMARY}
\medskip
{\footnotesize
\noindent
{\it Title of program}: {\susy23}
\par
\noindent
{\it Program obtainable from}:~CPC Program Library, Queen's University
of Belfast, N.Ire\-land (see application form in this issue) and from
ftp.kek.jp in directory kek/minami/susy23.
\par
\noindent
{\it Computer for which the program is designed and others on which it is 
operable}:~HP9000 and most of the UNIX platforms with a FORTRAN77 compiler 
\par
\noindent
{\it Computer}: HP9000 ; 
\par
{\it Installation}: 
High Energy Accelerator Research Organization(KEK), 
Tsuku\-ba, Ibaraki, Japan
\par
\noindent
{\it Operating system}:~UNIX ; 
\noindent
\par
{\it Programming language used}: FORTRAN77
\par
\noindent
{\it High speed storage required}: 23 Mbyte ; 
\noindent
{\it Card image code}: ASCII
\par
\noindent
{\it Key words}: supersymmetry, $e^+e^-$ colliders, 
ISR, QEDPS, event generator, spin correlation,
hadronization.
\par
}
\end{minipage}
~~~
\begin{minipage}{7.6cm}
\medskip
{\footnotesize
\noindent
{\it Nature of physical problem}
\par
\noindent
Study of supersymmetric particle search at LEP2.
\par
\noindent
{\it Method of solution}
\par
\noindent
The automatic amplitude generator {\tt GRACE} is used to get the
necessary helicity amplitudes 
for twenty-three sparticle production processes.
The specific corrections such as 
the initial state radiation or hadronization
are implemented in the program.
Sparticle widths and decay branching ratios 
for some modes are calculated.
Each event of final state is then generated 
according to these probabilities.
\par
\noindent
{\it Typical running time}
\par
\noindent
The running time depends on the number of diagrams of the
selected process, on the required cross-section accuracy and
on the applied cuts. 
For instance, on a HP-755/99, the process 
$e^+e^- \to e^+{\tilde{e}}_R{\chi}^0_1$ takes 20
minutes to reach a 0.5\% accuracy
on the total cross section.
\par
\noindent
}
\end{minipage}

%% file: chap12.tex
\medskip
\leftline{\large\bf 1. Introduction}
\smallskip

The supersymmetric (SUSY) standard model \cite{Nilles} is the most
promising extension of the standard model (SM)
as it could naturally give a solution
to the gauge hierarchy problem. 
In SUSY model, thanks to the symmetry between bosons and fermions, 
the quadratic divergences are genuinely cancelled out.
The price to pay is the existence of a large number of new particles,
SUSY particles (sparticles), 
yet to be discovered.
The most impressive evidence in favor of SUSY may be
the unification of gauge couplings in SUSY Grand Unified 
Theories (GUTs) \cite{running}. 
The Lightest SUSY Particle (LSP) is often proposed as a candidate 
to the missing mass in the Universe for the cold component of the 
so-called dark matter \cite{dm}. 
Many peoples consider the "SUSY world" as a plausible scenario 
for future particle physics prompting the need of event generators.

Although the gauge hierarchy issue set a mass upper bound to SUSY 
particles of $O$(1TeV), lighter sparticles are seen as giving more 
natural solution.
So there is a possibility that the sparticles could 
be discovered at LEP2 in electron-positron collision.  
While the hadron colliders such as FNAL Tevatron and DESY HERA 
have already covered 
larger scattering energies, LEP2 could be the first machine 
to discover ``light'' sparticles (the chargino, 
the slepton or the stop) as the background conditions are far better in 
electron-positron than in hadron colliders.

{\susy23} is an event generator suitable for 
detector simulation and data analysis dedicated to
SUSY processes in the LEP2 energy range.
 
It is based on the {\tt GRACE} \cite{grace,gracesusy} system
which generates 
automatically the matrix element in terms of helicity amplitudes 
(supplied by the {\tt CHANEL} \cite{chanel} library) for 
any processes once the initial and the final states have been specified. 
In addition, 
a kinematics library has been developed 
for each process topology 
(weeding out multidimensional singularities) for a better
convergence of the Monte-Carlo integration over the phase 
space.

The {\susy23} package is actually a collection of 23 types of 
SUSY processes at $e^+e^-$ colliders, 
consisting of 18 final 2-body and 5 final 3-body 
processes, presented in a coherent and 
uniform environment. 
These processes are
\medskip

 $e^+e^- \to$

\medskip

\begin{tabular}{p{5em}p{5em}p{5em}p{5em}p{5em}p{5em}}
{\mbox{$\chi^{+}_1$}}{\mbox{$\chi^{-}_1$}} & 
{\mbox{$\chi^{0}_1$}}{\mbox{$\chi^{0}_2$}} & 
{\mbox{$\chi^{0}_2$}}{\mbox{$\chi^{0}_2$}} &
{\mbox{$\tilde{\bar{\nu}}_e$}}{\mbox{$\tilde{\nu}_e$}}     & 
{\mbox{$\tilde{\bar{\nu}}_\mu$}}{\mbox{$\tilde{\nu}_\mu$}} &
{\mbox{$\tilde{\bar{\nu}}_{\tau}$}}{\mbox{$\tilde{\nu}_{\tau}$}}\\ 
{\mbox{$\tilde{\tau}^{+}_1$}}{\mbox{$\tilde{\tau}^{-}_1$}} & 
{\mbox{$\tilde{\tau}^{+}_2$}}{\mbox{$\tilde{\tau}^{-}_2$}} &
{\mbox{$\tilde{\tau}^{+}_1$}}{\mbox{$\tilde{\tau}^{-}_2$}} &
{\mbox{$\tilde{\tau}^{+}_2$}}{\mbox{$\tilde{\tau}^{-}_1$}} &
{\mbox{$\tilde{e}^{+}_L$}}{\mbox{$\tilde{e}^{-}_L$}} & 
{\mbox{$\tilde{e}^{+}_R$}}{\mbox{$\tilde{e}^{-}_R$}} \\ 
{\mbox{$\tilde{e}^{+}_L$}}{\mbox{$\tilde{e}^{-}_R$}} &
{\mbox{$\tilde{e}^{+}_R$}}{\mbox{$\tilde{e}^{-}_L$}} &
{\mbox{$\tilde{\mu}^{+}_L$}}{\mbox{$\tilde{\mu}^{-}_L$}} & 
{\mbox{$\tilde{\mu}^{+}_R$}}{\mbox{$\tilde{\mu}^{-}_R$}} &
{\mbox{$\tilde{\bar{t}}_1$}}{\mbox{$\tilde{t}_1$}} &
{\mbox{$\tilde{\bar{b}}_1$}}{\mbox{$\tilde{b}_1$}} \\ 
\end{tabular}

\medskip
and

\medskip
\begin{tabular}{p{5em}p{5em}p{5em}p{5em}p{5em}p{5em}}
{\mbox{$e^{+}$}}{\mbox{$\tilde{e}^{-}_R$}}{\mbox{$\chi^{0}_1$}}&
{\mbox{$e^{+}$}}{\mbox{$\tilde{e}^{-}_L$}}{\mbox{$\chi^{0}_1$}} &
{\mbox{$e^{+}$}}{\mbox{$\tilde{\nu}_e$}}{\mbox{$\chi^{-}_1$}} &
{\mbox{$\nu_e$}}{\mbox{$\tilde{e}^{+}_L$}}{\mbox{$\chi^{-}_1$}}&
{\mbox{$\gamma$}}{\mbox{$\chi^{0}_1$}}{\mbox{$\chi^{0}_1$}} & \\
\end{tabular}

\bigskip
Once a process has been selected, the total and differential cross 
sections are computed
with the Monte-Carlo integration package {\tt BASES} \cite{bases}.
Then {\tt SPRING} \cite{bases}, a general purpose event generator, 
provides unweighted events.
Some physics results have already been presented in 
ref.\cite{ours}. 

The matrix element generated by {\tt GRACE} corresponds
to the genuine tree level process in the massive case. 
Initial state radiation corrections, 
unstable sparticle decays and hadronization  
must then be introduced to produce realistic event generation.
All of them have been implemented in {\susy23}.

For the initial state radiation (ISR) two techniques are provided 
in the program.
\footnote{ISR is not available for the hard photon process, i.e.,
{\mbox{$e^{-}e^{+} \rightarrow \gamma \chi^{0}_1 \chi^{0}_1$.}}} 
The first one uses the well-known analytic form of $e^\pm$ structure 
function \cite{kuraev} 
and the second is based on {\tt QEDPS} \cite{isr}, a radiative correction
generator
producing an indefinite numbers of photons according to the parton 
shower algorithm in the leading-logarithmic (LL) approximation. 
Originally this
algorithm has been developed to simulate QCD parton shower.
One important point here is that {\tt QEDPS} reproduces
the radiative photon transverse momentum distributions. 

Sparticle widths and decay branching ratios 
are calculated 
(see Tables 3$\sim$8 in Appendix C). 
Each final state is then generated according to
the branching ratios. 
In the present version, 2-body and 3-body direct decays as well as 
some possible cascade decays are included. 
Note that helicity informations of matrix elements 
of the direct decays of inos 
\footnote{We use ino as a generic name to represent neutral (neutralinos) 
and charged (charginos) mixtures of gauginos and higgsinos.} 
are taken into account in evaluating their decays.

We assume that the hadronization of partons can be separated
from the hard interaction studied here.
Under this assumption, the calculation of cross sections is
exact in {\susy23}. 
Final state parton hadronization is performed, in {\susy23},
through the  mechanism  implemented in 
{\tt JETSET} \cite{jetset}. 
 

This paper is organized as follows. 
MSSM basics is given in section~2.
The structure of the program is discussed in 
section~3. 
In section 4, all details about
running the program are presented. 
A summary of the paper can be found in section 5. 
Four appendices describe the parameters and options 
which can be changed by the user, the list of all processes and the 
program installation procedure. 

\medskip
\medskip
\medskip
\medskip
\leftline{\large\bf 2. Theoretical framework of MSSM} 
\smallskip

\medskip
\leftline{\large\bf 2.1 Particle content}
\smallskip

The Minimal SUSY Standard Model (MSSM)  includes the {\it minimal} particle 
content.
That is, there should be at least one new particle (sparticle) for each 
known particle in the SM and the two Higgs doublets. 
The additional Higgs doublet must be included in order to 
give mass to the up- and down-type quarks and to allow for the chiral 
anomaly cancellation \cite{Nilles}. 
 
For quarks and charged leptons there exist two scalar 
partners per species, while for neutrinos there exists
one scalar partner per species.
These scalars are called squarks and sleptons, or, generically, sfermions. 
Before SU(2)$\times$U(1) breaking, the left- and right-sfermions 
do not mix since they have different SU(2)$\times$U(1) quantum numbers. 
After the breaking, however, they can mix with each other. 
In this case the mass eigenstates are parametrized by a mixing angle $\theta_f$ ; 
\begin{equation}
\left({\sf_1\atop\sf_2}\right)=
\left(
{\sf_{L}\,\cos\theta_f+\sf_{R}\,\sin\theta_f}
\atop
{-\sf_{L}\,\sin\theta_f+\sf_{R}\,\cos\theta_f}
\right). 
\end{equation}
Actually this mixing effect is substantial only for the third-generation 
sfermions, especially for the stops (superpartners of the top quark) 
owing to the large top mass \cite{stop}. 

The new fermions are either the superpartner of spin-1 gauge bosons 
(gauginos) or that of spin-0 bosons (higgsinos). 
They mix each other when SU(2)$\times$U(1) symmetry is broken.
The mass eigenstates (inos) are usually mixtures of gaugino and 
higgsino states.
They are called neutralinos 
{\mbox{$\chi^0_k$}} ($k=1\sim4$) 
and charginos {\mbox{$\chi^{\pm}_i$}} ($i=1,2$), respectively, 
according to their electric charges.  
Gluinos $\gluino$ are free from mixing 
since the color SU(3) is not broken.
As usual, we consider the lightest neutralino $\chi^0_1$ as 
the lightest sparticle (LSP).

MSSM contains two Higgs doublets and 
five physical Higgs bosons 
are left after SU(2)$\times$U(1) breaking \cite{higgs}.
They are two CP even ($h^0$, $H^0$), 
one CP odd ($A^0$) neutral scalars and 
two charged scalars ($H^\pm$). 

The list of sparticles in the model is shown in Appendix A, 
where the abbreviated particle names used in the control card and 
the KF-codes assigned for each sparticles in {\susy23} are also
 shown. 

\medskip
\medskip
\medskip
\leftline{\large\bf 2.2 Basic parameters of MSSM}
\smallskip

The standard model has 18 fundamental parameters to be determined 
by experiments. 
The MSSM has a somewhat larger number of parameters. 
They are classified as 
({\romannumeral 1}) gauge couplings, 
({\romannumeral 2}) superpotential parameters and 
({\romannumeral 3}) soft-breaking parameters. 
In the following, we adopt the notation of the SUSY parameters described 
in ref.\cite{Hikasa}.

\medskip
\leftline{\large\bf 2.2.1 gauge couplings}
\smallskip

The three gauge coupling parameters corresponding to  
SU(3), SU(2) and U(1) gauge groups are the same as in the SM.  
All gauge interactions are governed by these couplings. 
They determine the fermion-sfermion-gaugino interactions 
and four-point scalar interactions as well as ordinary 
fermion-fermion-gauge-boson interactions. 

\medskip
\leftline{\large\bf 2.2.2 superpotential parameters and $R$-parity}
\smallskip

In the MSSM, ordinary Yukawa interactions are generalized using the 
superpotential $W(\hat{\phi})$, where $\hat{\phi}$ denotes arbitrary 
chiral superfield \cite{Nilles}. 
Renormalizability restricts the functional form of the superpotential to 
\begin{equation}
W(\hat{\phi})=m_{ij}\hat{\phi}_{i}\hat{\phi}_{j}
+\lambda_{ijk}\hat{\phi}_i\hat{\phi}_j\hat{\phi}_k.
\end{equation}
The parameters $m_{ij}$ and $\lambda_{ijk}$ are further constrained by 
the gauge symmetry and some discrete symmetries. 

Later, we use the notation $\mu$ which is
a coefficient of the term quadratic in Higgs superfields
and contributes to
the mass terms of higgsinos and Higgs bosons \cite{Hikasa}.

A  well-known discrete, multiplicative symmetry is the $R$-parity  
defined by  
\begin{equation}
R=(-)^{3(B-L)+2S},
\end{equation}  
where $B$, $L$ and $S$ stand for the baryon number,
the lepton number and the spin, respectively. 
This formula implies that all ordinary SM particles 
have even $R$-parity, whereas the corresponding superpartners 
have odd $R$-parity.  
Usually we impose the $B-L$ conservation on the MSSM 
and then the MSSM possesses the $R$-parity invariance. 
In {\susy23} only $R$-parity conserving processes are
considered.

\medskip
\leftline{\large\bf 2.2.3 soft-breaking parameters}
\smallskip

SUSY is not an exact symmetry of nature 
since our world is not manifestly supersymmetric. 
In the MSSM the SUSY breaking is induced by the soft-SUSY breaking 
terms, which do not introduce quadratic divergences. 
Hence the solution of the naturalness problem remains intact. 
There are four types of soft breaking terms ; 
(1)  gaugino masses $M_i$  ($i=1\sim3$), 
(2)  masses for the sfermions $\mtil_{f}$,  
(3)  trilinear term  $A_f$ and 
(4)  three Higgs mass terms. 
 The three mass parameters in the last item
 can be re-expressed in terms of 
 two Higgs vacuum expectation values, $v_1$ and $v_2$, and 
 one physical Higgs mass. 
 Here $v_1$ and $v_2$ respectively denote the vacuum expectation values 
 of the Higgs field coupled to $d$-type and $u$-type quarks. 
 $v_1^2 + v_2^2$ $=$ (246 GeV)$^2$ is determined from 
 the experimentally measured $W$-boson mass, 
 while the ratio 
 \begin{equation}
 \tanbe = {\frac{v_2}{v_1}}
 \end{equation}
 is a free parameter of the model. 
 
 A generally accepted assumption is that 
 all the three gaugino mass parameters $M_i$ are equal at some 
 grand unification scale $M_X$. 
 Then the gaugino mass parameters can be expressed in terms of 
 one of them, for instance, $M_2$. 
 The other two gaugino mass parameters are given by 
 \begin{eqnarray}
 M_1 &=& {\frac{5}{3}}\tan^2{\tew} M_2 \\
 M_3 &=& {\frac{\alpha_3}{\alpha}}\sw M_2,  
 \end{eqnarray}
where $\alpha$ and $\alpha_3$ ($=\alpha_s$) denote the QED and the QCD 
coupling constants, respectively. 

The trilinear terms $A_f$ appear in off-diagonal elements of the 
mass matrix for sfermions($\widetilde{f}_L$, $\widetilde{f}_R$). 
Consequently, it has direct relation to the mass eigenvalues
($m_{\widetilde{f}_1}$, $m_{\widetilde{f}_2}$)  and 
the mixing angle $\theta_f$. 
In our formulation, $A_f$'s ($f=t, b, \tau$) are
determined by those mixing parameters (see Eqs.(19), (20) and (21)).

\medskip
\medskip
\medskip
\leftline{{\large\bf 2.3 MSSM parameters in} {\susy23}}
\smallskip


In Appendix B, the input MSSM parameters in {\susy23} are listed. 
They are 
\begin{itemize}
\item[i)]  gaugino parameters, ($\tanbe$, $M_2$, $\mu$), 
\item[ii)] masses of scalar partners of charged leptons 
in the 1st and 2nd generations, 
($\msel$, $\mser$ ), ($\msml$, $\msmr$ ),

\item[iii)] masses of 1st and 2nd generation squarks, 
($\msul$, $\msur$, $\msdr$), ($\mscl$, $\mscr$  $\mssr$)

\item[iv)] masses and mixing angles of charged sfermions 
in the 3rd generation, 
($\mstl$, $\msth$, $\tht$), ($\msbl$, $\msbh$, $\thb$),
($\mstaul$, $\tha$) and $\msnt$. 

\end{itemize}
\medskip

From the above inputs, 
the following physical parameters are calculated in 
{\susy23}, 
\begin{itemize}
\item[i)] masses of {\mbox{$\chi^0_i$}} ($i=1\sim4$), the
mixing matrix {\mbox{${\cal O}_N$}} and
the sign factor $\eta_i$ of neutralinos,

\begin{eqnarray}
{\mbox{$\chi^0_{iL}$}} &=& \eta_i ({\cal O}_N)_{ij} {\cal X}^0_{jL}, \\
{\mbox{$\chi^0_{iR}$}} &=& \eta_i^{*} ({\cal O}_N)_{ij} {\cal X}^0_{jR},
\end{eqnarray}

with ${\cal X}^0_{iL} = ( {\widetilde{B}}_L, 
{\widetilde{W}}^0_L , {\widetilde{H}}^0_{1L}, 
{\widetilde{H}}^0_{2L}) $.
\item[ii)] masses of {\mbox{$\chi^\pm_i$}}($i=1\sim2$), 
the mixing angles {\mbox{$\cos\phi_{L}$}}, {\mbox{$\cos\phi_{R}$}}, 
{\mbox{$\sin\phi_{L}$}}, {\mbox{$\sin\phi_{R}$}}, 
sign factor {\mbox{$\epsilon_R$}} of charginos,

\begin{eqnarray}
{\mbox{$\chi^-_{1L}$}} &=& 
{\widetilde{W}}^-_L \cos\phi_L + 
{\widetilde{H}}^-_{1L} \sin\phi_L, \\
{\mbox{$\chi^-_{2L}$}} &=& 
- {\widetilde{W}}^-_L \sin\phi_L + 
{\widetilde{H}}^-_{1L} \cos\phi_L, \\
{\mbox{$\chi^-_{1R}$}} &=& {\widetilde{W}}^-_R \cos\phi_R
+ {\widetilde{H}}^-_{2L} \sin\phi_R, \\
{\mbox{$\chi^-_{2R}$}} &=& \epsilon_R 
( - {\widetilde{W}}^-_R \sin\phi_R + 
{\widetilde{H}}^-_{2L} \cos\phi_R ),
\end{eqnarray}

\item[iii)] the gluino mass $\msg$ ($=M_3$) using Eq.(6), 

\item[iv)] $\msne$ and $\msnm$ are calculated by the following relations:
\begin{eqnarray}
\msne^{2} &=& 
\msel^{2}  -m_{e}^{2} + m_{W}^{2}\cos2{\beta}, \\
\msnm^{2} &=& 
\msml^{2}  -m_{\mu}^{2} + m_{W}^{2}\cos2{\beta}. 
\end{eqnarray}

\item[v)] $\msdl$ and $\mssquarkl$ are calculated by the following relations:
\begin{eqnarray}
\msdl^{2} &=& \msul^{2}  -m_{u}^{2} +m_{d}^{2} - m_{W}^{2}\cos2{\beta},
  \\
\mssquarkl^{2} &=&
\mscl^{2}  -m_{c}^{2}
  + m_{s}^{2} - m_{W}^{2}\cos2{\beta},  
\end{eqnarray}

\item[vi)] $\mstauh$  is given by:
\begin{eqnarray}
\msnt^{2} &=& 
\cos^2{\tha}\mstaul^{2}+ \sin^2{\tha}\mstauh^{2} -m_{\tau}^{2}
+ m_{W}^{2}\cos2{\beta}, 
\end{eqnarray}

\item[vii)] $\msbh$  is given by:
\begin{eqnarray}
\cos^2{\tht}\mstl^{2}+ \sin^2{\tht}\msth^{2} -m_{t}^{2} &=&
\cos^2{\thb}\msbl^{2}+ \sin^2{\thb}\msbh^{2} -m_{b}^{2}
+ m_{W}^{2}\cos2{\beta}, 
\end{eqnarray}

\item[viii)] the trilinear term  $A_f$ using formulae as follows 
\begin{eqnarray}
A_t&=&-{\frac{1}{2}}\sin{2\theta_t}{\frac{m^2_{\st_2}-m^2_{\st_1}}
{m_t}}-\mu\cot{\beta} \label{tpara} \\
A_b&=&-{\frac{1}{2}}\sin{2\theta_b}{\frac{m^2_{\sb_2}-m^2_{\sb_1}}
{m_b}}-\mu\tan{\beta}, \label{bpara} \\
A_\tau&=&-{\frac{1}{2}}\sin{2\theta_\tau}{\frac{m^2_{\stau_2}-m^2_{\stau_1}}
{m_\tau}}-\mu\tan{\beta}, \label{apara} 
\end{eqnarray}


\end{itemize}

Here we briefly discuss the relation between 
our input parameters and those adopted in the 
{\tt susygen} program \cite{susygen}.
As for the gaugino parameters, the sign convention of $\tan\beta$ and 
$\mu$ is different. We use the positive definite $\mu$ and 
$\tan\beta$ can have either positive or negative values. 
In  {\tt susygen}, on the other hand, positive definite $\tan\beta$ 
and positive or negative $\mu$ are adopted. 
As there exits a symmetry for the sign of $\tan\beta$ and $\mu$ 
in the MSSM, 
the negative $\mu$ in {\tt susygen} corresponds to the negative 
$\tan\beta$ in {\susy23}.
The definition of gaugino mass $M_2$ is the same in both generators. 
As for the left-right mixing of sfermions in the 3rd generation, 
we take observable masses and mixing angles, 
($m_{\widetilde{f}_1}$, $m_{\widetilde{f}_2}$, $\theta_f$), 
as inputs. 
In {\tt susygen}, on the other hand, more basic model parameters 
($m_{\widetilde{f}_L}$, $m_{\widetilde{f}_R}$, $A_f$) are adopted. 
The relation between them can be found from 
Eqs.(\ref{tpara}) $\sim$ (\ref{apara}) and the following formula, 
\begin{equation}
m^{2}_{{\widetilde{f}_1}\atop{\widetilde{f}_2}}
         ={\frac{1}{2}}\left[ m^{2}_{\widetilde{f}_{L}}+m^{2}_{\widetilde{f}_{R}}
             \mp \left( (m^{2}_{\widetilde{f}_{L}}-m^{2}_{\widetilde{f}_{R}})^{2}
            +(2a_{f}m_{f})^{2}\right)^{1/2}\right],
\label{sfmass}
\end{equation}
where $a_t = -(A_t + \mu\cot\beta)$, $a_b = -(A_b + \mu\tan\beta)$ and 
$a_{\tau} = -(A_{\tau} + \mu\tan\beta)$. 
We did not impose low energy mass relations between masses of sfermions 
determined by the renormalization group equations in some SUSY GUTs 
because such mass relations sensitively depend on details of a model and 
boundary conditions for model parameters at the unification scale.  
In {\tt susygen}, however, there is an option for the calculation in the  
framework of the minimal supergravity model \cite{Hikasa}. 
In this model, masses of all sfermions in the 1st and 2nd generations are 
determined by an input parameter, $m_\infty$ ($= m_0$), 
which is the universal soft scalar mass at the unification scale. 
For convenience of the users, 
we present an example of relations between observable masses 
and the parameter $m_\infty$ 
for the case of $\sin^2\theta_W$ $=$ $0.230$, 
\begin{eqnarray}
\msul^2 &=& m^2_\infty + 7.26M^2_\infty + m^2_u
     + ({\frac{1}{2}}-{\frac{2}{3}}\sin^2\theta_W) m^2_Z \cos{2\beta}, \\
\msur^2 &=& m^2_\infty + 6.83M^2_\infty + m^2_u
     + {\frac{2}{3}}\sin^2\theta_W m^2_Z \cos{2\beta}, \\
\msdr^2 &=& m^2_\infty + 6.78M^2_\infty + m^2_d
     - {\frac{1}{3}}\sin^2\theta_W m^2_Z \cos{2\beta}, \\
\msel^2 &=& m^2_\infty + 0.530M^2_\infty + m^2_e
     - ({\frac{1}{2}}-\sin^2\theta_W) m^2_Z \cos{2\beta}, \\
\mser^2 &=& m^2_\infty + 0.1515M^2_\infty + m^2_e
     - \sin^2\theta_W m^2_Z \cos{2\beta}, 
\end{eqnarray}
where $M_\infty$ $=$ $1.22 M_2$ denotes the universal soft gaugino mass 
at the unification scale.

%% file: chap3.tex
\medskip
\leftline{\large\bf 3. Features of the program}
\smallskip
This section
covers some important features of the program, the introduction of the initial
state radiation, available decay modes of unstable sparticles and
the spin correlation in the ino decays.

\medskip

\medskip
\leftline{\large\bf 3.1 Initial state radiation}
\smallskip
In the first approach, the simple electron structure function
is used.
The electron structure function at
$O(\alpha^2)$~\cite{kuraev}  which is to be
convoluted with the cross section for a primary
process is given by
\begin{eqnarray}
  D(x,s)&=& \left[1+{3\over8}\beta
 +\left({9\over128}-{\zeta(2)\over8}\right)\beta^2\right]
                 {\beta\over2}(1-x)^{\beta/2-1} \nonumber \\
    && -{\beta\over 4} \bigl( 1+x \bigr)
        -{\beta^2 \over 32} \biggl[ 4(1+x)\ln(1-x)
              +{1+3x^2 \over 1-x}\ln x+(5+x) \biggr],
                    \label{eq:Dxs} \\
\beta&=&(2\alpha/\pi)(\ln(s/m_e^2)-1),
\end{eqnarray}
where $s$ is the square of the total energy of the system and $x$ is
the momentum fraction of the electron.
Compared with the exact $O(\alpha)$ calculation for 
$s$-channel annihilation,
the corrected cross section obtained by this function
does not contain the so-called $K$-factor:
\begin{equation}
1+{\alpha\over\pi}\left({\pi^2\over3}-{1\over2}\right)=1.006480\cdots.
\label{eq:kfact}
\end{equation}
This factor is not included since it is sensible only when the 
$s$-channel $e^+e^-$ annihilation takes place.
The basic assumption of the QED Parton Shower method, {\tt QEDPS} 
\cite{isr}, 
is primarily based on the fact that $D(x,Q^2)$ obeys the 
Altarelli-Parisi equation, which can be expressed by the integral
equation
in the leading-logarithmic(LL) approximation
\begin{equation}
D(x,Q^2)= \Pi(Q^2,Q_s^2)D(x,Q_s^2)
+{\alpha\over2\pi}\int\nolimits_{Q_s^2}^{Q^2}{dK^2\over K^2}
    \Pi(Q^2,K^2)\int\nolimits_x^{1-\epsilon}{dy\over y}
           P(y)D(x/y,K^2),       \label{eq:intform}
\end{equation}
where the small quantity $\epsilon$ is a cutoff related with the  
infrared singularity \cite{ll} and it will be defined later.
In this equation $P(x)$ is the split function noted $P_+(x)$ 
when regularized at $x=1$.
$Q_s^2$ is the initial value of $Q^2$ and a free parameter(of order
$m_e^2$). 
For simplicity the fine structure constant $\alpha$ is assumed not 
running with $Q^2$. 
The Sudakov factor $\Pi$ is given by:
\begin{equation}
 \Pi(Q^2,{Q'}^2) = \exp\left(- {\alpha\over 2 \pi} \int_{{Q'}^2}^{Q^2}
 { d K^2 \over K^2} \int_0^{1-\epsilon} d x P(x) \right).
  \label{eq:non}
\end{equation}
and denotes
the probability that an electron evolves from ${Q'}^2$ to 
$Q^2$ without emitting hard photon. 

The integral equation (\ref{eq:intform}) can be solved by iteration.
It is clear that 
the emission of $n$ photons corresponds to $n$ iterations. Hence 
it is possible to regard the process as a stochastic mechanism 
suggesting the shower algorithm in Ref.~\cite{isr}.

Once an exclusive process is fixed by the algorithm, each branching of
a photon is a real process, that is, 
an electron with $x,K^2$ decays as:
    $ e^-(x,-K^2)\to e^-(xy,-{K'}^2)+\gamma(x(1-y),Q_0^2)$.
Here $Q_0^2$ is a cut-off to avoid the infrared divergence 
and $\epsilon$ is given by $\epsilon = Q_0^2 / {K'}^2$. 
The momentum
conservation at the branching gives:
$ -K^2=-{K'}^2/y+Q_0^2/(1-y)+{\bf k}_T^2/(y(1-y)) $
which in turn determines the photon transverse momentum relative to
the parent, ${\bf k}_T^2$, from $y,K^2,{K'}^2$. 
This technique gives the ${\bf k}_T^2$ distribution as well as
the shape of the $x$-distribution.

The above algorithm concerns the 
case where either $e^-$ or $e^+$ radiates photons 
when the axial gauge vector is chosen along the momentum of the other 
electron, namely $e^+$ or $e^-$. In the program, however, we 
use the double cascade scheme to 
ensure the symmetry of the radiation between $e^+$ and 
$e^-$~\cite{double}. These two are mathematically
equivalent in the LL approximation. 

\medskip
\leftline{\large\bf 3.2 Sparticle decays}
\smallskip
\medskip
\leftline{\large\bf 3.2.1 Decay modes}
\smallskip
The decay modes shown below are implemented in \susy23.
They are also presented in Tables 3\(\sim\) 9 in Appendix C.
The decay modes (1)\(\sim\)(8) are applied iteratively
as long as the mass relationship between the sparticle and their 
decay products allows. 
The spin correlation is considered as is discussed in the
next subsection.
\begin{description}
\item[(1)] lighter chargino (Table 3)
\begin{eqnarray*}
{\chi^{-}_1}   &\to& f {\bar{f'}} {\chi^{0}_1}\\
        &\to& \ell {\bar{\tilde{\nu}}_{\ell}} \to \ell \quad ({\bar{\nu}_{\ell}} {
\chi^{0}_1}) \\
        &\to& {\bar{\nu}_{\ell}} {\tilde{\ell}} \to {\bar{\nu}_{\ell}} \quad
(\ell {\chi^{0}_1}) \\
        &\to& b {\bar{\tilde{t}}_1} \to b \quad (\bar{c} {\chi^{0}_1})
\end{eqnarray*}
\item[(2)] second lightest neutralino (Table 4)
\begin{eqnarray*}
{\chi^{0}_2}   &\to&  f \bar{f} {\chi^{0}_1}\\
               &\to&  \ell {\tilde{\ell}} \to \ell \quad (\ell {\chi^{0}_1})\\
               &\to&  \nu {\tilde{\nu}} \to \nu \quad (\nu {\chi^{0}_1})
\end{eqnarray*}
\item[(3)] left-handed selectron and smuon (Table 5)
\begin{eqnarray*}
{\tilde{\ell}_L} &\to& \ell {\chi^{0}_1}\\
         &\to& \ell {\chi^{0}_2} \to \ell \quad (f\bar{f} {\chi^{0}_1})\\
         &\to& \nu {\chi^{-}_1} \to \nu \quad (f\bar{f'}{\chi^{0}_1})
\end{eqnarray*}
\item[(4)] right-handed selectron and smuon (Table 6)
\begin{eqnarray*}
{\tilde{\ell}_R} &\to& \ell {\chi^{0}_1}\\
         &\to& \ell {\chi^{0}_2} \to \ell \quad (f\bar{f} {\chi^{0}_1})
\end{eqnarray*}
\item[(5)] lighter and heavier staus (Table 7)
\begin{eqnarray*}
{\tilde{\tau}_{1,2}} &\to& \tau {\chi^{0}_1}\\
         &\to& \tau {\chi^{0}_2} \to \tau \quad (f\bar{f} {\chi^{0}_1})\\
         &\to& \nu_{\tau} {\chi^{-}_1} \to \nu_{\tau} \quad (f\bar{f'}{\chi^
{0}_1})
\end{eqnarray*}
\item[(6)] sneutrinos (Table 8)
\begin{eqnarray*}
{\tilde{\nu}_{\ell}} &\to& \nu_{\ell} {\chi^{0}_1}\\
         &\to& \nu_{\ell} {\chi^{0}_2} \to \nu_{\ell} \quad (f\bar{f} {\chi^
{0}_1})\\
         &\to& \ell^+ {\chi^{-}_1} \to \ell^+ \quad (f\bar{f'}{\chi^{0}_1})
\end{eqnarray*}
\item[(7)] lighter stop (Table 9)
\begin{eqnarray*}
{\tilde{t}_1} &\to& c {\chi^{0}_1}\\
         &\to& c {\chi^{0}_2} \to c \quad (f\bar{f} {\chi^{0}_1})\\
         &\to& b {\chi^{-}_1} \to b \quad (f\bar{f'}{\chi^{0}_1})
\end{eqnarray*}
\item[(8)] lighter sbottom (Table 9)
\begin{eqnarray*}
{\tilde{b}_1} &\to& b {\chi^{0}_1}\\
         &\to& b {\chi^{0}_2} \to b \quad (f\bar{f} {\chi^{0}_1})
\end{eqnarray*}
\end{description}
We assume that squarks
in the first and second generations and the gluinos are sufficiently
heavy to forbid the charginos, neutralinos, sleptons,
stop and sbottom decays in these channels. 
These are natural assumptions
as long as one deal with production processes of
the sparticles with mass $\nle 100$GeV in
the LEP2 energy regions.

In \susy23 system, the masses are inspected  whether
the decay is possible or not. 
\footnote{
Users must be careful in taking the input mass parameters so that 
the lightest neutralino $\chi^0_1$ (LSP) should be lighter than 
charginos and sfermions.}
When the channel is open, the decay width is computed.

\medskip
\leftline{\large\bf 3.2.2 Spin correlations}
\smallskip
The helicity informations of matrix elements in the fermionic
2-body and 3-body decays of the lighter chargino ${\chi^{-}_1}$ and
the second lightest neutralino ${\chi^{0}_2}$ are used
in the event generation of their decay products. In other words 
kinematical distribution of the decay products is sensitive to
spin correlations with the mother ino.
All decay matrix elements are computed exactly
with {\tt GRACE} \cite{grace} in tree approximation.

In the event generation step, 
the helicity state of inos is
selected 
by the Monte-Carlo method  based on the
relative cross sections of all possible helicity combinations.
Four-momentum vectors of daughter-particle are
determined by the decay matrix element of the 
ino with specified helicity.

%% file: chap4.tex
\medskip
\leftline{\large\bf  4. Structure of the program}
\smallskip

The {\susy23} provides event generation for 
twenty-three SUSY processes as listed in Table 2 in Appendix C,
at $e^+e^-$ colliders  with or without radiative corrections.
As is shown in Appendix B,
there are many options covering theoretical and experimental
requirements.
Since all program components are  distributed as source code, 
users can select all options
by editing the relevant subprograms directly.
However, 
an interface program {\tt susy23} is prepared
to lighten the user's burden. 
\footnote{
Please do not confuse the executable module named
{\tt susy23} with the \susy23 system itself.
}
It selects and/or corrects the program code which are 
affected by the various
options and it creates a ``{\tt Makefile}'' according to the user 
requirements.
This procedure is called the source generation phase.

In the integration step, the matrix element of a selected
process is integrated over the phase space 
by the subprogram {\tt BASES},
which gives the total and differential cross sections and
the probability distribution
used in the event generation phase \cite{bases}.
There, the subprogram {\tt SPRING} samples a point
in the phase space and test if it can be accepted as a new event
according to its probability.
When an event is accepted, the program control returns 
to the main program, where sparticle decays are performed.

There are, therefore, three steps in the generator {\susy23}, 
the first is
the source generation, the second is the integration and the third is
the event generation.
In addition to the user interface program {\tt susy23}, the following
programs are available in the {\susy23} system:
\begin{itemize}
\item[i)] Twenty-three function programs {\tt FUNC}s for final 3-body
or 2-body SUSY processes, each of which calculates the numerical
value of the differential cross section for each process.
\item[ii)] 
The kinematics subprogram {\tt KINMOQ} is called when initial state
radiation (ISR) based on QEDPS, the QED parton shower model, is
requested. Although {\tt KINEMO} covers both the case where no
ISR is produced and where at most 2 photons can be generated
at zero angle following the electron structure function method.
\item[iii)] An extended version of the {\tt CHANEL} \cite{chanel}
library
to include MSSM couplings. The {\tt CHANEL} library
calculates the numerical values of Feynman 
diagrams based on the helicity amplitudes method.
\item[iv)] The numerical integration and event generation program
package {\tt BASES/SPRING v5.1} \cite{bases}.
\item[v)] The main programs, {\tt MAINBS} and {\tt MAINSP}, and
all program components for the integration and event generation steps.
\item[vi)] The subprograms for branching ratios and sparticle decays.
\item[v)] The interface programs to {\tt JETSET} 
 ({\tt GR2LND}, {\tt SP2LND} and {\tt GRC2SH}).
\end{itemize}
\par\noindent
The function of these program components and relationship among them 
are presented in the next three subsections.
\par

\medskip
\medskip
\medskip
\leftline{\large\bf 4.1 Source generation step}
\smallskip
The user interface program {\tt susy23} reads the parameters from
the control data file, which contains
process selection, type of radiative corrections,
SUSY parameters, etc.
The complete list of parameters is found in
Appendix B.
Then {\tt susy23} generates the following program code:

\par
\begin{itemize}
\item[i)] Three initialization subprograms {\tt USRPRM}, {\tt MODMAS}
          {\tt KINIT}, {\tt USERSP} and
\item[ii)] A "{\tt Makefile}".
\end{itemize}

\medskip
\leftline{\large\bf 4.2 Integration step}
\smallskip
\par
Before starting the numerical integration, the main program {\tt MAINBS}
invokes an initialization subprogram {\tt USERIN}, in which the
following subprograms are called in this order:
\par
\smallskip
\begin{tabular}{llp{30em}}
{\tt USRPRM} & : & To define the set of input SUSY parameters 
                  and some optional parameters. \\
{\tt SETMAS} & : & To set default values to particle masses and 
decay widths. \\
{\tt MODMAS} & : & To alter the default values of all parameters defined
in {\tt SETMAS}.\\
{\tt AMPARM} & : & To set the coupling constants and other 
parameters.\\
{\tt KINIT}  & : & To set the parameters for the integration,
kinematics, cuts and to initialize the histogram package.
\end{tabular}
\par
The subprograms {\tt SETMAS} and {\tt AMPARM} are generated
by the {\tt GRACE} system.
By default ino and charged Higgs masses, as well as ino 
mixing angles (see Sec.2.3) and all branching ratios, are calculated
in the framework of the MSSM. 
Users can selectively set the values of these parameters
and some branching ratios.
Only naive consistency checks are made,
such as the normalization check of the branching ratios
and the mass ordering check, 
so all modifications on these parameters or branching ratios in
the subprograms {\tt USRPRM}, {\tt MODMAS} and {\tt KINIT} are
on the users' own responsibility.
\par
The integration program {\tt BASES} calculates the scattering
cross section by sampling the function {\tt FUNC} on the allowed
phase space segmented by a self adapted grid where finer
cells are clustered on the high gradient zones. This is an
iterative procedure
running until either the maximum number of allowed iteration
is reached or the required accuracy is obtained.
In the function program {\tt FUNC},
the kinematics subroutine {\tt KINEMO}
or {\tt KINMOQ} maps the integral variables
into the four-momentum of the final state particles.
{\tt KINEMO} is called for reactions with no radiative corrections
or those
involving the initial state radiation 
treated with the structure function
techniques. {\tt KINMOQ} is called for processes 
in which radiative
corrections are computed with the QED parton shower method.
The subprograms {\tt AMPTBL} and {\tt AMPSUM} are further called 
for calculating the helicity amplitudes and the 
squared of their sum.

{\bf It is recommended to look at the integration result carefully, 
especially over the convergency behaviors both 
for the grid optimization and integration steps.}
When the accuracy of each iteration fluctuates from iteration to iteration
or when 
it jumps up suddenly to a large value compared 
to the other
iterations, the resultant estimate of the integral may not be reliable.
There are two possible origins of this behavior; too few
sampling points or/and an unsuitable choice of
the kinematical variables.

After the numerical integration by {\tt BASES},
the subprograms {\tt BSINFO} 
and {\tt BHPLOT} are called to print the result of integration and
the histograms, respectively.
Before terminating the integration procedure the probability
distribution of the integrand can be saved in a file named
{\tt bases.data}
by invoking {\tt BSWRIT}, which is later used for the event generation
by {\tt SPRING}.
\par

\medskip
\medskip
\leftline{\large\bf 4.3 Event generation step}
\smallskip
\par
The main program, {\tt MAINSP}, handles
the event generation step.
The subprogram {\tt BSREAD} reads the data file {\tt bases.data}
to restore
the probability distribution and then the subprogram {\tt USERIN}
is called.
Each call to {\tt SPRING} generates one event by sampling a point
in the phase volume. It first 
calculates the differential cross section at that point using 
the same function {\tt FUNC} in the integration step and
returns the weight of this sampling point.
A weight one event is finally produced using usual unweighting
technique. 
Subsequently, {\tt SUDCAY} handles the sparticle decay.
By calling the subprogram {\tt SP2LND},
the event information  
including the color connection to be used by {\tt JETSET} 
is stored in the labelled common {\tt LUJETS}. 

Then hadronization of quarks and gluons can be
performed by calling {\tt LUEXEC}.
At the end of the event generation, the routine {\tt SPINFO} and 
{\tt SHPLOT}
are invoked successively for printing event generation information and
histograms.


%% file: howto.tex
\leftline{\large\bf 5. How to run the program}

The user should first prepare the control data to define the process, 
the option flags, the MSSM parameters and the experimental cuts.
The user interface program {\tt susy23} takes this control data 
as an input. For example, let's call the following sequence of lines
{\tt control.data}.
\par\smallskip
{\tt
\begin{tabular}{lll}
process & = & SW1sw1 \\
energy  & = & 200.0d0 \\
type    & = & tree \\
hadron  & = & no \\
tanbe  & = & -2.0d0 \\
xm2  & = & 50.0d0 \\
xmu  & = & 150.0d0 \\
end & &  
\end{tabular}
}
\par\smallskip\noindent
The first line specifies the process to be calculated and
the second is the center of mass energy in GeV unit.
The others are options, 
whose meanings are given in Appendix B.
Then the user may type:
\par
\begin{verbatim}
% susy23 < control.data 
\end{verbatim}
\par\noindent
If the message ``syntax error'' is returned,
the user must carefully examine the contents of the control data.
No file is generated in this case.
When the control data is accepted
the following message appears:
\par
\begin{verbatim}
Process is "SW1sw1"
Energy is "200.0d0"
HADRNZ <no>
bye-bye
directory name is SW1sw1
absolute directory name is /home/susy23/prc/SW1sw1
------------------------------------------
cd /home/susy23/prc/SW1sw1
make integ
integ
make spring
spring
------------------------------------------
\end{verbatim}
\par\noindent
According to the parameters given in the control data, 
the files, i.e.~ {\tt usrprm.f}, 
{\tt modmas.f}, {\tt kinit.f},
{\tt usersp.f} and {\tt Makefile}, are generated in
a specified subdirectory ( {\tt SW1sw1} in this case).
\par\noindent
According to the last five lines in the message, 
users can proceed with the calculations as follows:
\par
\begin{itemize}
\item[i)] Change directory by typing:
\begin{verbatim}
% cd /home/susy23/prc/SW1sw1
\end{verbatim}
\item[ii)] Create an executable {\tt integ} for the integration by typing:
\begin{verbatim}
% make integ
\end{verbatim}
\par
\item[iii)] Numerical integration is actually performed by typing:
\begin{verbatim}
%  integ
\end{verbatim}
\par
The results of
integration step are displayed on the console as well as written in an
output file 
{\tt bases.result}. The total cross section in $pb$ and the
estimated statistical error are shown 
on the last line, under {\tt Cumulative Result}, 
in the table of the {\tt Convergence Behavior for the Integration step}.
The differential cross sections are also printed as a function of the energy, 
scattering angle of each particle and invariant masses of any two final particles. 
The probability distribution of the integrand is written in a file 
{\tt bases.data} which will be used in the event generation step by {\tt spring}.

\item[iv)] Before running the event generation, users may edit 
{\tt mainsp.f} to set additional parameters if needed 
and call user's own analysis routines.

The following is the structure of the generated {\tt mainsp.f}, where
four-momentum of all particles are stored in the {\tt common/lujets/} in the 
{\tt JETSET} format when subprogram {\tt sp2lnd} is called in the event-loop:
\begin{verbatim}
      Program  mainsp
      implicit real*8(a-h,o-z)
      external func
       ....................
      real*4  p,v
      common/lujets/n,k(4000,5),p(4000,5),v(4000,5)
       ....................
       ....................
      mxtry  =   50
      mxevnt = 1000
      do 100 nevnt = 1, mxevnt

         call spring( func, mxtry )
       ....................
*         -----------------
           call sp2lnd
*         -----------------
*
*       ==============================================
*       (  user_analysis based on the common lujets )
*       ==============================================
*
  100 continue
       ....................
      stop
      end
\end{verbatim}

\item[v)] Create an executable {\tt spring} for event generation
by typing:

\begin{verbatim}
% make spring
\end{verbatim}

\item[vi)] Start the event generation by typing:

\begin{verbatim}
% spring
\end{verbatim}

Information on the event generation will be written in the
{\tt spring.result} file. Users should pay attention to the 
histograms generated in this step. 
The distributions of the generated events are superimposed
with the character ``{\tt 0}'' on the histograms generated in
the integration step. These two
distributions should be consistent with each other 
within the statistical error of
the generation. For the detail of the output files of {\tt BASES}
and {\tt SPRING},
users can consult the Ref.\cite{bases}. 
\end{itemize}

%% file: summary.tex
\leftline{\large\bf 6. Summary}


The {\susy23} system calculates the effective cross 
section and generates events for one of 
the twenty-three SUSY processes
at $e^+e^-$ 
colliders listed in Appendix C. 
It is dedicated to the LEP2 studies. 
The numbers 2 and 3 in \susy23 refer to the fact that 
processes with 2 and also 3 final particles are fully computed at
tree level in the massive case.
These particles are then allowed to decay producing 
multi-particle final states.
Sparticle widths and decay branching ratios 
in accordance with this energy range
are calculated.
In particular, the helicity information for 
direct decays of spinor sparticles (the lighter chargino and 
the second lightest neutralino) is handled properly.
The produced 
quarks can be hadronized according to {\tt JETSET}. 
Processes with initial radiations can be 
generated in terms of the electron structure function or the QED 
parton shower method. 




%% file: ack.tex
\leftline{\large\bf Acknowledgements}
The authors would like to thank G. Coignet, S. Rosier-Lees, 
F. Boudjema,  S. Komamiya,  S. Asai,  P. Bambade and R. Tanaka  
for their interest and encouragement and colleagues in Minami-Tateya
group of KEK for their help. This work was done in the KEK-LAPP
collaboration supported in part by Mombusho in Japan 
under the Grant-in-Aid for International Scientific Research Program 
No.07044097, and CNRS/IN2P3 in France.

%% file: ref.tex

%% file: appa.tex
\newcommand{\nue}{{\nu_e}}
\newcommand{\nul}{{\nu_\ell}}
\newcommand{\nuebar}{{\bar{\nu}_e}}
\newcommand{\numu}{{\nu_\mu}}
\newcommand{\numubar}{{\bar{\nu}_\mu}}
\newcommand{\nutau}{{\nu_\tau}}
\newcommand{\nutaubar}{{\bar{\nu}_\tau}}
\newcommand{\electron}{{e^-}}
\newcommand{\positron}{{e^+}}
\newcommand{\muon}{{\mu^-}}
\newcommand{\antimuon}{{\mu^+}}
\newcommand{\taum}{{\tau^-}}
\newcommand{\antitau}{{\tau^+}}
\newcommand{\uq}{{u}}
\newcommand{\ubar}{{\bar{u}}}
\newcommand{\dq}{{d}}
\newcommand{\dbar}{{\bar{d}}}
\newcommand{\cq}{{c}}
\newcommand{\cbar}{{\bar{c}}}
\newcommand{\sq}{{s}}
\newcommand{\sbar}{{\bar{s}}}
\newcommand{\bq}{{b}}
\newcommand{\bbar}{{\bar{b}}}
\def    \be             {\begin{equation}}
\def    \ee             {\end{equation}}
\def    \chipm          {\mbox{$\chi^{\pm}$}}
\def    \chipma        {\mbox{$\chi^{\pm}_1$}}
\def    \chipmb        {\mbox{$\chi^{\pm}_2$}}
\def    \chipa        {\mbox{$\chi^{+}_1$}}
\def    \chima        {\mbox{$\chi^{-}_1$}}
\def    \chipb        {\mbox{$\chi^{+}_2$}}
\def    \chimb        {\mbox{$\chi^{-}_2$}}
\def    \chio          {\mbox{$\chi^0$}}
\def    \chioa         {\mbox{$\chi^0_1$}}
\def    \chiob         {\mbox{$\chi^0_2$}}
\def    \chioc         {\mbox{$\chi^0_3$}}
\def    \chiod         {\mbox{$\chi^0_4$}}
\def    \stop           {\mbox{$\tilde{t}$}}
\def    \stopa           {\mbox{$\tilde{t}_1$}}
\def    \sbottoma           {\mbox{$\tilde{b}_1$}}
\def    \sbottomb           {\mbox{$\tilde{b}_2$}}
\def    \sbottomabar           {\mbox{$\tilde{\bar{b}}_1$}}
\def    \sbottombbar           {\mbox{$\tilde{\bar{b}}_2$}}
\def    \stopabar           {\mbox{$\tilde{\bar{t}}_1$}}
\def    \stopb           {\mbox{$\tilde{t}_2$}}
\def    \stopbbar           {\mbox{$\tilde{\bar{t}}_2$}}
\def    \sel            {\mbox{$\tilde{e}^-_L$}}
\def    \ser            {\mbox{$\tilde{e}^-_R$}}
\def    \selbar            {\mbox{$\tilde{e}^+_L$}}
\def    \serbar            {\mbox{$\tilde{e}^+_R$}}
\def    \sml            {\mbox{$\tilde{\mu}^-_L$}}
\def    \smr            {\mbox{$\tilde{\mu}^-_R$}}
\def    \smlbar            {\mbox{$\tilde{\mu}^+_L$}}
\def    \smrbar            {\mbox{$\tilde{\mu}^+_R$}}
\def    \staua            {\mbox{$\tilde{\tau}^-_1$}}
\def    \staub            {\mbox{$\tilde{\tau}^-_2$}}
\def    \stauabar            {\mbox{$\tilde{\tau}^+_1$}}
\def    \staubbar            {\mbox{$\tilde{\tau}^+_2$}}
\def    \sne            {\mbox{$\tilde{\nu}_e$}}
\def    \snl            {\mbox{$\tilde{\nu}_\ell$}}
\def    \snebar            {\mbox{$\tilde{\bar{\nu}}_e$}}
\def    \snmu            {\mbox{$\tilde{\nu}_\mu$}}
\def    \snmubar            {\mbox{$\tilde{\bar{\nu}}_\mu$}}
\def    \sntau            {\mbox{$\tilde{\nu}_\tau$}}
\def    \sntaubar            {\mbox{$\tilde{\bar{\nu}}_\tau$}}
\def    \sll            {\mbox{$\tilde{\ell}_L$}}
\def    \slr            {\mbox{$\tilde{\ell}_R$}}
\def    \sul            {\mbox{$\tilde{u}_L$}}
\def    \sur            {\mbox{$\tilde{u}_R$}}
\def    \sulbar            {\mbox{$\tilde{\bar{u}}_L$}}
\def    \surbar            {\mbox{$\tilde{\bar{u}}_R$}}
\def    \sdl            {\mbox{$\tilde{d}_L$}}
\def    \sdr            {\mbox{$\tilde{d}_R$}}
\def    \sdlbar            {\mbox{$\tilde{\bar{d}}_L$}}
\def    \sdrbar            {\mbox{$\tilde{\bar{d}}_R$}}
\def    \scl            {\mbox{$\tilde{c}_L$}}
\def    \scr            {\mbox{$\tilde{c}_R$}}
\def    \sclbar            {\mbox{$\tilde{\bar{c}}_L$}}
\def    \scrbar            {\mbox{$\tilde{\bar{c}}_R$}}
\def    \ssl            {\mbox{$\tilde{s}_L$}}
\def    \ssr            {\mbox{$\tilde{s}_R$}}
\def    \sslbar            {\mbox{$\tilde{\bar{s}}_L$}}
\def    \ssrbar            {\mbox{$\tilde{\bar{s}}_R$}}

\leftline{\large\bf Appendix A. List of sparticles in the MSSM}

\vspace{2mm}

The list of sparticles in the MSSM is shown in Table 1, which includes 
abbreviated names used in the control cards and the KF-codes 
asigned for each particles. 
As is seen, the keywords in \susy23 are case-sensitive.

\vspace{2mm}

\begin{center}
\begin{tabular}{|l|l|r||l|l|r|} \hline
particle & abbrev. & KF code &
particle & abbrev. & KF code \\ \hline \hline
\rule{0mm}{4mm}
$\chipa$ & {\tt SW1} & 75 &
$\sulbar$ & {\tt SUL}& $-$42 \\ \hline
\rule{0mm}{4mm}
$\chima$ & {\tt sw1}& $-$75 &
$\sur$   & {\tt sur }& 48  \\ \hline
\rule{0mm}{4mm}
$\chipb$ & {\tt SW2}& 76 &
$\surbar$ & {\tt SUR}& $-$48 \\ \hline
\rule{0mm}{4mm}
$\chimb$ & {\tt sw2}& $-$76& 
$\sdl$   & {\tt sdl}& 41 \\ \hline
\rule{0mm}{4mm}
$\chioa$ & {\tt sz1}& 71 &
$\sdlbar$ & {\tt SDL} & $-$41 \\ \hline
\rule{0mm}{4mm}
$\chiob$ & {\tt sz2}& 72 &
$\sdr$   & {\tt sdr}& 47 \\ \hline
\rule{0mm}{4mm}
$\chioc$ & {\tt sz3}& 73 &
$\sdrbar$ & {\tt SDR}& $-$47 \\ \hline
\rule{0mm}{4mm}
$\chiod$ & {\tt sz4}& 74 & 
$\scl$ & {\tt scl}& 44 \\ \hline
\rule{0mm}{4mm}
$\gluino$ & {\tt sgl}&  70 & 
$\sclbar$ & {\tt SCL}& $-$44 \\ \hline
\rule{0mm}{4mm}
$\sel$ & {\tt sel}& 51 &
$\scr$ & {\tt scr}& 50 \\ \hline
\rule{0mm}{4mm}
$\selbar$ & {\tt SEL}& $-$51 & 
$\scrbar$ & {\tt SCR}& $-$50 \\ \hline
\rule{0mm}{4mm}
$\ser$ & {\tt ser}& 57 &
$\ssl$ & {\tt ssl}& 43 \\ \hline
\rule{0mm}{4mm}
$\serbar$ & {\tt SER}& $-$57 &
$\sslbar$ & {\tt SSL}& $-$43 \\ \hline
\rule{0mm}{4mm}
$\sne$ & {\tt sne}& 52 & 
$\ssr$ & {\tt ssr}& 49 \\ \hline
\rule{0mm}{4mm}
$\snebar$ & {\tt SNE}& $-$52 &
$\ssrbar$ & {\tt SSR}& $-$49 \\ \hline
\rule{0mm}{4mm}
$\sml$ & {\tt sml}& 53 &
$\stopa$ & {\tt st1}& 46 \\ \hline
\rule{0mm}{4mm}
$\smlbar$ & {\tt SML}& $-$53 &
$\stopabar$ & {\tt ST1}& $-$46 \\ \hline
\rule{0mm}{4mm}
$\smr$ & {\tt smr}& 58 &
$\stopb$ & {\tt st2}& 62 \\ \hline
\rule{0mm}{4mm}
$\smrbar$ & {\tt SMR}& $-$58 & 
$\stopbbar$ & {\tt ST2}& $-$62 \\ \hline
\rule{0mm}{4mm}
$\snmu$ & {\tt snm}& 54 &
$\sbottoma$ & {\tt sb1}& 45 \\ \hline
\rule{0mm}{4mm}
$\snmubar$ & {\tt SNM}& $-$54 &
$\sbottomabar$ & {\tt SB1}& $-$45 \\ \hline
\rule{0mm}{4mm}
$\staua$ & {\tt sa1}& 55 & 
$\sbottomb$ & {\tt sb2}& 61 \\ \hline
\rule{0mm}{4mm}
$\stauabar$ & {\tt SA1}& $-$55 & 
$\sbottombbar$ & {\tt SB2}& $-$61 \\ \hline
\rule{0mm}{4mm}
$\staub$ & {\tt sa2}& 59 & 
$h^0$ & {\tt sh1}& 25 \\ \hline
\rule{0mm}{4mm}
$\staubbar$ & {\tt SA2}& $-$59 &
$H^0$ & {\tt sh2}& 35 \\ \hline
\rule{0mm}{4mm}
$\sntau$ & {\tt sna}& 56 &
$A^0$ & {\tt sh3}& 36 \\ \hline
\rule{0mm}{4mm}
$\sntaubar$ & {\tt SNA}& $-$56 & 
$H^+$ & {\tt sh4}& 37 \\ \hline
\rule{0mm}{4mm}
$\sul$ & {\tt sul}& 42 & 
$H^-$ & {\tt SH4}& $-$37 \\ \hline
\end{tabular}
\end{center}

\begin{center}
Table 1 \ Sparticles in \susy23.
\end{center}

%% file: appb.tex
\leftline{\large\bf Appendix B. Parameters in control data }

In the table below, the default values are underlined 
and  the relation between commands and variable/array in
Fortran sources is also described. Variable names are written in
bold letters and filenames in italic. 
All masses and energy parameters are given in GeV.

\begin{itemize}
\item[i)] Process selection.

\begin{tabular}{|p{4.2em}cp{25em}|} \hline
{\tt Process}&{\tt =}& \underline{\tt SW1sw1} \\
 & & abbreviation of process name \\ \hline
\end{tabular}

This specifies the subdirectory name, where 
subroutines for the process are stored.
Table 1 in Appendix C shows the abbreviation
of process names.

\item[ii)] Center of mass energy.

\begin{tabular}{|p{4.2em}cp{25em}|} \hline
{\tt energy}&{\tt =}& \underline{\tt 200.D0} \\
 & & CMS energy \\ 
& &{\tt w} in {\it kinit.f} \\ \hline
\end{tabular}

\item[iii)] Control parameters.

\begin{tabular}{|p{4.2em}cp{25em}|} \hline
{\tt helicity1}   &{\tt =}& {\tt \underline{average}},
~{\tt left}, ~{\tt right}\\
 & & Helicity state for the initial electron.\\
{\tt helicity2}   &{\tt =}& {\tt \underline{average}},
~{\tt left}, ~{\tt right}\\
 & & Helicity state for the initial positron.\\
\hline
\end{tabular}

\medskip

\begin{tabular}{|p{4.2em}cp{25em}|} \hline
{\tt type}   &{\tt =}& {\tt \underline{tree}}, ~{\tt sf}, 
~{\tt qedpsi} \quad
(Not available for $\gamma{\widetilde{\chi}}^0_1{\widetilde{\chi}}^0_1$)\\
& & Type of calculation: \\
 & & Without radiation({\tt tree}), 
ISR with structure function({\tt sf}) and 
ISR with QEDPS({\tt qedpsi}). \\
& & {\tt jqedps} in {\it usrprm.f}, {\tt isr} 
in {\it kinit.f}. \\
& & {\tt jqedps = 0}, {\tt isr = 0} without radiation({\tt tree}).\\
& & {\tt jqedps = 0}, {\tt isr = 1} for {\tt sf}.\\
& & {\tt jqedps = 1} for {\tt qedpsi}.\\
 \hline
\end{tabular}

\medskip

\begin{tabular}{|p{4.2em}cp{25em}|} \hline
{\tt decay}   &{\tt =}& {\tt \underline{yes}}, ~{\tt no} \\
& & Unstable sparticle decays :\\
& & {\tt jdecay = \underline{1}} or {\tt 0} in {\it usrprm.f}.\\ \hline
%
%
{\tt spincr}   &{\tt =}& {\tt yes}, ~{\tt \underline{no}} \\
& & Include spin correlations for direct 3-body decays 
of gauginos : \\
& & {\tt jspncr = 1} or {\tt \underline{0}} in {\it usrprm.f}.\\ \hline
{\tt qcdcr}   &{\tt =}& {\tt yes}, ~\underline{\tt no} \\
& & Include overall QCD correction factor in squark production : \\
& & {\tt jqcdcr = 1} or {\tt \underline{0}} in {\it usrprm.f}.\\ \hline
{\tt susyp}   &{\tt =}& {\tt yes}, ~{\tt \underline{no}} \\
& & (yes) \ user can selectively set ino masses and mixing angles as well as 
charged Higgs mass in {\it modmas.f} (see next sections) \\
& & (no) \ all parameters are calculated in the MSSM framework : \\
& & {\tt jsusyp = 1} or {\tt \underline{0}} in {\it usrprm.f}.\\ \hline
\end{tabular}

\item[iv)] Standard Model parameters.

\begin{tabular}{|p{4.2em}cp{25em}|} \hline
{\tt amw}   &{\tt =}& {\tt \underline{80.23D0}} \\
 & & $W$ mass : {\tt amw} in {\it modmas.f}.\\ \hline
{\tt agw}   &{\tt =}& {\tt \underline{2.08D0}} \\
 & & $W$ width : {\tt agw} in {\it modmas.f}.\\ \hline
{\tt amz}   &{\tt =}& {\tt \underline{91.19D0}} \\
 & & $Z$ mass : {\tt amz} in {\it modmas.f}.\\ \hline
{\tt agz}   &{\tt =}& {\tt \underline{2.491D0}} \\
 & & $Z$ width : {\tt agz} in {\it modmas.f}.\\ \hline
{\tt amu}   &{\tt =}& {\tt \underline{100.0D-3}} \\
 & & u-quark mass : {\tt amu} in {\it modmas.f}.\\ \hline
{\tt amd}   &{\tt =}& {\tt \underline{100.0D-3}} \\
 & & d-quark mass : {\tt amd} in {\it modmas.f}.\\ \hline
{\tt amc}   &{\tt =}& {\tt \underline{1.8645D0}} \\
 & & c-quark mass : {\tt amc} in {\it modmas.f}.\\ \hline
{\tt ams}   &{\tt =}& {\tt \underline{300.0D-3}} \\
 & & s-quark mass : {\tt ams} in {\it modmas.f}.\\ \hline
{\tt amt}   &{\tt =}& {\tt \underline{174.0D0}} \\
 & & top mass : {\tt amt} in {\it modmas.f}.\\ \hline
{\tt amb}   &{\tt =}& {\tt \underline{5.0D0}} \\
 & & b-quark mass : {\tt amb} in {\it modmas.f}.\\ \hline
\end{tabular}

\medskip

\begin{tabular}{|p{4.2em}cp{25em}|} \hline
{\tt alphai}   &=& {\tt \underline{128.07D0}} \\
 & & $\alpha^{-1}$ : {\tt alphai} in {\it modmas.f}.\\ \hline
{\tt alpha\_s}   &=& {\tt \underline{0.12D0}} \\
 & & $\alpha_s$ : {\tt alphas} in {\it modmas.f}.\\ \hline
\end{tabular}


\item[v)] Minimal SUSY Standard Model parameters.

\begin{tabular}{|p{4.2em}cp{25em}|} \hline
{\tt tanbe}   &{\tt =}& {\tt \underline{-2.0D0}} \\
 & & $\tan\beta$ : {\tt tanbe} in {\it usrprm.f}.\\ \hline
{\tt xm2}   &{\tt =}& {\tt \underline{50.0D0}} \\
 & & SU(2) gaugino mass $M_2$ : {\tt xm2} in {\it usrprm.f}.\\ \hline
{\tt xmu}   &{\tt =}& {\tt \underline{150.0D0}} \\
 & & SUSY Higgs mixing mass $\mu$ : {\tt xmu} in {\it usrprm.f}.\\ \hline
\end{tabular}

\newpage

\item[vi)] Sfermion sector parameters.

\begin{tabular}{|p{4.2em}cp{25em}|} \hline
{\tt xmsel}   &{\tt =}& {\tt \underline{95.0D0}} \\
 & & left-handed selectron mass $\msel$ : 
{\tt xmsel} in {\it usrprm.f}. \\ \hline
{\tt xmser}   &{\tt =}& {\tt \underline{90.0D0}} \\
 & & right-handed selectron mass $\mser$ :
 {\tt xmser} in {\it usrprm.f}.\\ \hline
{\tt xmsml}   &{\tt =}& {\tt \underline{95.0D0}} \\
 & & left-handed smuon mass $\msml$ : 
{\tt xmsml} in {\it usrprm.f}. \\ \hline
{\tt xmsmr}   &{\tt =}& {\tt \underline{90.0D0}} \\
 & & right-handed smuon mass $\msmr$ :
 {\tt xmsmr} in {\it usrprm.f}.\\ \hline
\end{tabular}

\medskip

\begin{tabular}{|p{4.2em}cp{25em}|} \hline
{\tt xmsa1}   &{\tt =}& {\tt \underline{90.0D0}} \\
 & & lighter stau mass $\mstaul$ : 
{\tt xmsa1} in {\it usrprm.f}.\\ \hline
{\tt xmsa2}   &{\tt =}& {\tt \underline{95.0D0}} \\
 & & heavier stau mass $\mstauh$ : 
{\tt xmsa2} in {\it usrprm.f}. (this parameter is recalculated
in case of {\tt susyp = no})
\\ \hline
{\tt xmsna}   &{\tt =}& {\tt \underline{100.0D0}} \\
 & & tau-sneutrino mass $\msnt$ :
 {\tt xmsna} in {\it usrprm.f}.\\ \hline
{\tt tha}   &{\tt =}& {\tt \underline{pi/2.0D0}} \\
 & & mixing angle $\tha$ of staus :
 {\tt tha} in {\it usrprm.f}.\\ \hline
\end{tabular}

\medskip

\begin{tabular}{|p{4.2em}cp{25em}|} \hline
{\tt xmsul}   &{\tt =}& {\tt \underline{200.0D0}} \\
 & & left-handed s u-quark mass $\msul$ :
 {\tt xmsul} in {\it usrprm.f}. \\ \hline
{\tt xmsur}   &{\tt =}& {\tt \underline{200.0D0}} \\
 & & right-handed s u-quark mass $\msur$ :
 {\tt xmsur} in {\it usrprm.f}.\\ \hline
{\tt xmsdr}   &{\tt =}& {\tt \underline{200.0D0}} \\
 & & right-handed s d-quark mass $\msdr$ : 
{\tt xmsdr} in {\it usrprm.f}.\\ \hline
{\tt xmscl}   &{\tt =}& {\tt \underline{200.0D0}} \\
 & & left-handed s c-quark mass $\mscl$ :
 {\tt xmscl} in {\it usrprm.f}. \\ \hline
{\tt xmscr}   &{\tt =}& {\tt \underline{200.0D0}} \\
 & & right-handed s c-quark mass $\mscr$ :
 {\tt xmscr} in {\it usrprm.f}.\\ \hline
{\tt xmssr}   &{\tt =}& {\tt \underline{200.0D0}} \\
 & & right-handed s s-quark mass $\mssr$ :
 {\tt xmssr} in {\it usrprm.f}.\\ \hline
\end{tabular} 

\medskip

\begin{tabular}{|p{4.2em}cp{25em}|} \hline
{\tt xmst1}   &{\tt =}& {\tt \underline{80.0D0}} \\
 & & lighter stop mass $\mstl$ : 
{\tt xmst1} in {\it usrprm.f}.\\ \hline
{\tt xmst2}   &{\tt =}& {\tt \underline{250.0D0}} \\
 & & heavier stop mass $\msth$ :
 {\tt xmst2} in {\it usrprm.f}.\\ \hline
{\tt xmsb1}   &{\tt =}& {\tt \underline{90.0D0}} \\
 & & lighter sbottom mass $\msbl$ : 
{\tt xmsb1} in {\it usrprm.f}.\\ \hline
{\tt xmsb2}   &{\tt =}& {\tt \underline{210.0D0}} \\
 & & heavier sbottom mass $\msbh$ :
 {\tt xmsb2} in {\it usrprm.f}.
This parameter is calculated
in case of {\tt susyp = no}.
\\ \hline
{\tt tht}   &{\tt =}& {\tt \underline{1.5D0}} \\
 & & mixing angle $\tht$ of stops : {\tt tht} in {\it usrprm.f}.\\ \hline
{\tt thb}   &{\tt =}& {\tt \underline{0.3D0}} \\
 & & mixing angle $\thb$ of sbottoms : {\tt thb} in {\it usrprm.f}.\\ \hline
\end{tabular}

\medskip

\begin{tabular}{|p{4.2em}cp{25em}|} \hline
{\tt xaa} &{\tt =}& {\tt \underline{0.0D0}} \\
& & trilinear coupling $A_\tau$ in {\it modmas.f} 
in case of {\tt susyp = yes}. \\ \hline
{\tt xau} &{\tt =}& {\tt \underline{250.0D0}} \\
& & trilinear coupling $A_t$ in {\it modmas.f} 
in case of {\tt susyp = yes}. \\ \hline
{\tt xad} &{\tt =}& {\tt \underline{50.0D0}} \\
& & trilinear coupling $A_b$ in {\it modmas.f} 
in case of {\tt susyp = yes}. \\ \hline
\end{tabular}

\newpage
\item[vii)] Neutralino sector parameters.

For only {\tt susyp = yes}, 
the following parameters can be given by users.

If {\tt susyp = no} is specified, the input for the
following parameters are neglected since these are
calculated by Eqs.(7) and (8).

\begin{tabular}{|p{4.2em}cp{25em}|} \hline
{\tt amn1} &{\tt =}& {\tt \underline{30.0D0}} \\
& & neutralino mass $\chi^0_1$.\\ \hline
{\tt amn2} &{\tt =}& {\tt \underline{70.0D0}} \\
& & neutralino mass $\chi^0_2$.\\ \hline
{\tt amn3} &{\tt =}& {\tt \underline{200.0D0}} \\
& & neutralino mass $\chi^0_3$.\\ \hline
{\tt amn4} &{\tt =}& {\tt \underline{400.0D0}} \\
& & neutralino mass $\chi^0_4$.\\ \hline
{\tt on} &{\tt =}& {\tt i},{\tt i}, {\tt \underline{1.0D0}} 
             (for {\tt i}$ = 1\sim4$)\\
             &{\tt =}&  {\tt i},{\tt j},
{\tt \underline{0.0D0}} (for {\tt i}$\neq${\tt j}, 
{\tt i},{\tt j} $= 1\sim4$)\\
& & neutralino mixing matrix ${\cal O}_N$ in {\it modmas.f} \\ \hline
{\tt sgn } &{\tt =}& {\tt i},
{\tt \underline{1.0D0}}, {\tt \underline{0.0D0}}
(for {\tt i} $= 1\sim4$)\\
& & neutralino sign factor $Re(\eta_i)$, $Im(\eta_i)$.\\
 \hline
\end{tabular}

\medskip

\item[viii)] Chargino sector parameters.

For only {\tt susyp = yes}, the following parameters can be
given by users.

If {\tt susyp = no} is specified, the input for the
following parameters are neglected since these are
calculated by Eqs.(9) $\sim$ (12).

\begin{tabular}{|p{4.2em}cp{25em}|} \hline
{\tt amc1} &{\tt =}& {\tt \underline{30.0D0}} \\
& & chargino mass $\chi^-_1$.\\ \hline
{\tt amc2} &{\tt =}& {\tt \underline{200.0D0}} \\
& & chargino mass $\chi^-_2$.\\ \hline
{\tt cphl} &{\tt =}& {\tt \underline{1.0D0}} \\
& & chargino mixing matrix $\cos\phi_L$.\\ \hline
{\tt sphl} &{\tt =}& {\tt \underline{0.0D0}} \\
& & chargino mixing matrix $\sin\phi_L$.\\ \hline
{\tt cphr} &{\tt =}& {\tt \underline{1.0D0}} \\
& & chargino mixing matrix $\cos\phi_R$.\\ \hline
{\tt sphr} &{\tt =}& {\tt \underline{0.0D0}} \\
& & chargino mixing matrix $\sin\phi_R$.\\ \hline
{\tt eps} &{\tt =}& {\tt \underline{1.0D0}} \\
& & chargino sign factor $\epsilon_R$.\\ \hline
\end{tabular}


\medskip

\item[ix)] Experimental cuts.

The numbering convention of particles  follows the {\tt GRACE} scheme,
where
the initial electron and positron are 1st and 2nd, respectively, and
the final particles are numbered 3, 4 and 5.
In the process name of Table 2,
the order of particles corresponds to this numbering convention.
For instance, in the process,
{\mbox{$e^{+}$}}{\mbox{$\tilde{e}^{-}_R$}}{\mbox{$\chi^{0}_1$}} ,
the 3rd is $e^+$, the 4th is $\tilde{e}^{-}_R$ and 
the 5th is $\chi^{0}_1$.

\begin{tabular}{|p{4.2em}cp{25em}|} \hline
{\tt thecut3}   &{\tt =}& {\tt \underline{180.D0}, \underline{0.0D0}} \\
 & & Polar angle cut for 3rd particle in degree 
(backward-angle, forward-angle). \\
 & & {\tt coscut(1:2,1)=cos(thecut3)} in {\it kinit.f}.\\ \hline
{\tt thecut4}   &{\tt =}& {\tt \underline{180.D0}, \underline{0.0D0}} \\
 & & Polar angle cut for 4th particle in degree 
(backward-angle, forward-angle). \\
 & & {\tt coscut(1:2,2)=cos(thecut4)} in {\it kinit.f}.\\ \hline
{\tt thecut5}   &{\tt =}& {\tt \underline{180.D0}, \underline{0.0D0}} \\
 & & Polar angle cut for 5th particle in degree 
(backward-angle, forward-angle). \\
 & & {\tt coscut(1:2,3)=cos(thecut5)} in {\it kinit.f}.\\ \hline
\end{tabular}

Instead of giving a numerical value, the user can use the 
strings as below:

{\tt amass1(i)} has the mass for {\it i}-th particle and
{\tt w} is the CM energy.

\begin{tabular}{|p{4.2em}cp{25em}|} \hline
{\tt engcut3}   &{\tt =}& {\tt \underline{amass1(3)}, \underline{w}} \\
 & & Energy cut for 3rd particle (min.,max.) \\
 & & {\tt engyct(1:2,1)} in {\it kinit.f} \\ \hline
{\tt engcut4}   &{\tt =}& {\tt \underline{amass1(4)}, \underline{w}} \\
 & & Energy cut for 4th particle (min.,max.) \\
 & & {\tt engyct(1:2,2)} in {\it kinit.f} \\ \hline
{\tt engcut5}   &{\tt =}& {\tt \underline{amass1(5)}, \underline{w}} \\
 & & Energy cut for 5th particle (min.,max.) \\
 & & {\tt engyct(1:2,3)} in {\it kinit.f} \\ \hline
\end{tabular}


\item[x)] Parameters for integration.

\begin{tabular}{|p{4.2em}cp{25em}|} \hline
{\tt itmx}   &{\tt =}& {\tt \underline{5}},~{\tt \underline{10}} \\
 & & Iteration numbers: {\tt itmx1}, {\tt itmx2} in {\it kinit.f} \\
 \hline
{\tt acc}   &=& {\tt \underline{0.1}}, ~{\tt \underline{0.05}} \\
 & & Accuracies in \%:{\tt acc1}, {\tt acc2}
 in {\it kinit.f} \\ \hline
{\tt ncall}   &=& {\tt \underline{10000}} \\
 & & Sampling points: {\tt ncall}
 in {\it kinit.f} \\ \hline
\end{tabular}

\item[xi)] Parameters for event generation.

\begin{tabular}{|p{4.2em}cp{25em}|} \hline
{\tt mxtry}   &=& \underline{{\tt 50}} \\
& & Maximum trial numbers: {\tt mxtry}
 in {\it usersp.f}\\ \hline
{\tt mxevnt}   &=& \underline{{\tt 10000}} \\
& & Maximum event numbers: {\tt mxevnt}
 in {\it usersp.f}\\ \hline
{\tt hadron}   &=& \underline{{\tt yes}}, {\tt no} \\
& & Hadronization with {\tt JETSET}\cite{jetset} for event generation:\\
& & If {\tt yes}, then a statement, {\tt call luexec} 
 is activate in {\it grc2sh.f}. \\ \hline
\end{tabular}

%
%
%

\item[xii)] End of description.

\begin{tabular}{|p{4.2em}cp{25em}|} \hline
{\tt end}   && \\ \hline
\end{tabular}

After the command {\tt end} any command is neglected.
\end{itemize}

%% file: appc.tex
\def    \be             {\begin{equation}}
\def    \ee             {\end{equation}}
\def    \chipm          {\mbox{$\chi^{\pm}$}}
\def    \chipma        {\mbox{$\chi^{\pm}_1$}}
\def    \chipmb        {\mbox{$\chi^{\pm}_2$}}
\def    \chipa        {\mbox{$\chi^{+}_1$}}
\def    \chima        {\mbox{$\chi^{-}_1$}}
\def    \chio          {\mbox{$\chi^0$}}
\def    \chioa         {\mbox{$\chi^0_1$}}
\def    \chiob         {\mbox{$\chi^0_2$}}
\def    \chioc         {\mbox{$\chi^0_3$}}
\def    \stop           {\mbox{$\tilde{t}$}}
\def    \stopa           {\mbox{$\tilde{t}_1$}}
\def    \sbottoma           {\mbox{$\tilde{b}_1$}}
\def    \stopabar           {\mbox{$\tilde{\bar{t}}_1$}}
\def    \stopb           {\mbox{$\tilde{t}_2$}}
\def    \sel            {\mbox{$\tilde{e}^-_L$}}
\def    \ser            {\mbox{$\tilde{e}^-_R$}}
\def    \selbar            {\mbox{$\tilde{e}^+_L$}}
\def    \serbar            {\mbox{$\tilde{e}^+_R$}}
\def    \sml            {\mbox{$\tilde{\mu}^-_L$}}
\def    \smr            {\mbox{$\tilde{\mu}^-_R$}}
\def    \smlbar            {\mbox{$\tilde{\mu}^+_L$}}
\def    \smrbar            {\mbox{$\tilde{\mu}^+_R$}}
\def    \stau            {\mbox{$\tilde{\tau}$}}
\def    \staul            {\mbox{$\tilde{\tau}^-_L$}}
\def    \staur            {\mbox{$\tilde{\tau}^-_R$}}
\def    \staulbar            {\mbox{$\tilde{\tau}^+_L$}}
\def    \staurbar            {\mbox{$\tilde{\tau}^+_R$}}
\def    \staua            {\mbox{$\tilde{\tau}^-_1$}}
\def    \staub            {\mbox{$\tilde{\tau}^-_2$}}
\def    \stauabar            {\mbox{$\tilde{\tau}^+_1$}}
\def    \staubbar            {\mbox{$\tilde{\tau}^+_2$}}
\def    \stauan            {\mbox{$\tilde{\tau}_1$}}
\def    \staubn            {\mbox{$\tilde{\tau}_2$}}
\def    \sne            {\mbox{$\tilde{\nu}_e$}}
\def    \snl            {\mbox{$\tilde{\nu}_\ell$}}
\def    \snebar            {\mbox{$\tilde{\bar{\nu}}_e$}}
\def    \snmu            {\mbox{$\tilde{\nu}_\mu$}}
\def    \snmubar            {\mbox{$\tilde{\bar{\nu}}_\mu$}}
\def    \sntau            {\mbox{$\tilde{\nu}_\tau$}}
\def    \sntaubar            {\mbox{$\tilde{\bar{\nu}}_\tau$}}
\def    \sll            {\mbox{$\tilde{\ell}_L$}}
\def    \slr            {\mbox{$\tilde{\ell}_R$}}

\leftline{\large\bf Appendix C. Process table}

\vspace{2mm}

The 23 processes included in \susy23 are listed in Table 2.
In the heading, 'abbrev.' and 'dir.' stand for the
abbreviated name used in the control card and directory 
name where the generated code is stored, respectively.
In Tables 3 $\sim$ 9, decay modes taken into account in the 
unstable sparticle decays in \susy23 are listed. 
By default, each branching ratio is automatically calculated 
in the framework of the MSSM. 
The abbreviated names are used in the control card or {\it kinit.f} 
when users want to set each branching ratio ({\tt jbrnch = 1}). 

\vspace{2mm}

\begin{center}
\begin{tabular}{|l|l|l||l|l|l|} \hline
process & abbrev. & dir. &
process & abbrev. & dir. \\ 
$e^+e^- \to$ &    &      &
$e^+e^- \to$ &    &      \\
\hline \hline
\rule{0mm}{4mm}
{\mbox{$\chi^{+}_1$}}{\mbox{$\chi^{-}_1$}} & {\tt SW1sw1} & {\tt SW1sw1} & 
{\mbox{$\tilde{\tau}^{+}_2$}}{\mbox{$\tilde{\tau}^{-}_1$}} & {\tt SA2sa1} & {\tt SA2sa1} \\ 
\hline 
\rule{0mm}{4mm}
{\mbox{$\chi^{0}_1$}}{\mbox{$\chi^{0}_2$}} & {\tt sz1sz2} & {\tt sz1sz2} & 
{\mbox{$\tilde{\tau}^{+}_2$}}{\mbox{$\tilde{\tau}^{-}_2$}} & {\tt SA2sa2} & {\tt SA2sa2} \\ 
\hline 
\rule{0mm}{4mm}
{\mbox{$\chi^{0}_2$}}{\mbox{$\chi^{0}_2$}} & {\tt sz2sz2} & {\tt sz2sz2} & 
{\mbox{$\tilde{\bar{\nu}}_e$}}{\mbox{$\tilde{\nu}_e$}}  & {\tt SNEsne} & {\tt SNEsne} \\ 
\hline 
\rule{0mm}{4mm}
{\mbox{$\gamma$}}{\mbox{$\chi^{0}_1$}}{\mbox{$\chi^{0}_1$}} & {\tt asz1sz1} & {\tt sz1sz1a} & 
{\mbox{$\tilde{\bar{\nu}}_\mu$}}{\mbox{$\tilde{\nu}_\mu$}} & {\tt SNMsnm} & {\tt SNMsnm} \\ \hline 
\rule{0mm}{4mm}
{\mbox{$\tilde{e}^{+}_R$}}{\mbox{$\tilde{e}^{-}_R$}} & {\tt SERser} & {\tt SERser} & 
{\mbox{$\tilde{\bar{\nu}}_{\tau}$}}{\mbox{$\tilde{\nu}_{\tau}$}}  & {\tt SNTsnt} & 
{\tt SNTsnt} \\ \hline 
\rule{0mm}{4mm}
{\mbox{$\tilde{e}^{+}_L$}}{\mbox{$\tilde{e}^{-}_L$}} & {\tt SELsel} & {\tt SELsel} & 
{\mbox{$e^{+}$}}{\mbox{$\tilde{e}^{-}_R$}}{\mbox{$\chi^{0}_1$}}  
& {\tt Esersz1} & {\tt Esersz1} \\ \hline 
\rule{0mm}{4mm}
{\mbox{$\tilde{e}^{+}_L$}}{\mbox{$\tilde{e}^{-}_R$}} & {\tt SELser} & {\tt SELser} & 
{\mbox{$e^{+}$}}{\mbox{$\tilde{e}^{-}_L$}}{\mbox{$\chi^{0}_1$}} & {\tt Eselsz1} & {\tt Eselsz1}  \\ \hline 
\rule{0mm}{4mm}
{\mbox{$\tilde{e}^{+}_R$}}{\mbox{$\tilde{e}^{-}_L$}} & {\tt SERsel} & {\tt SERsel} & 
{\mbox{$e^{+}$}}{\mbox{$\tilde{\nu}_e$}}{\mbox{$\chi^{-}_1$}} & {\tt Esnesw1} & 
{\tt Esnesw1} \\ \hline 
\rule{0mm}{4mm}
{\mbox{$\tilde{\mu}^{+}_R$}}{\mbox{$\tilde{\mu}^{-}_R$}} & {\tt SMRsmr} & {\tt SMRsmr} & 
{\mbox{$\nu_e$}}{\mbox{$\tilde{e}^{+}_L$}}{\mbox{$\chi^{-}_1$}} & {\tt nSELsw1} & {\tt nSELsw1}  \\ 
\hline 
\rule{0mm}{4mm}
{\mbox{$\tilde{\mu}^{+}_L$}}{\mbox{$\tilde{\mu}^{-}_L$}} & {\tt SMLsml} & {\tt SMLsml} & 
{\mbox{$\tilde{\bar{t}}_1$}}{\mbox{$\tilde{t}_1$}} & {\tt ST1st1} & {\tt ST1st1}  \\ 
\hline 
\rule{0mm}{4mm}
{\mbox{$\tilde{\tau}^{+}_1$}}{\mbox{$\tilde{\tau}^{-}_1$}} & {\tt SA1sa1} & {\tt SA1sa1} & 
{\mbox{$\tilde{\bar{b}}_1$}}{\mbox{$\tilde{b}_1$}} & {\tt SB1sb1} & {\tt SB1sb1}  \\ 
\hline 
\rule{0mm}{4mm}
{\mbox{$\tilde{\tau}^{+}_1$}}{\mbox{$\tilde{\tau}^{-}_2$}} & {\tt SA1sa2} & {\tt SA1sa2} &  &  & \\ 
\hline 
\end{tabular}
\end{center}

\begin{center}
Table 2 \ Available processes in \susy23.
\end{center}

\vspace{3mm}

\begin{center}
\begin{tabular}{|l|l||l|l|} \hline
process & abbrev. &
process & abbrev. \\ \hline \hline
\rule{0mm}{4mm}
$\chima\to\electron\nuebar\chioa$ & {\tt sw1br(1)} & 
$\chima\to\taum\sntaubar$ & {\tt sw1br(8)}  \\ \hline 
\rule{0mm}{4mm}
$\chima\to\muon\numubar\chioa$ & {\tt sw1br(2)} & 
$\chima\to\nuebar\sel$ & {\tt sw1br(9)}  \\ \hline 
\rule{0mm}{4mm}
$\chima\to\taum\nutaubar\chioa$ & {\tt sw1br(3)} & 
$\chima\to\numubar\sml$ & {\tt sw1br(10)}  \\ \hline 
\rule{0mm}{4mm}
$\chima\to\ubar\dq\chioa$ & {\tt sw1br(4)} & 
$\chima\to\nutaubar\staua$ & {\tt sw1br(11)}  \\ \hline 
\rule{0mm}{4mm}
$\chima\to\cbar\sq\chioa$ & {\tt sw1br(5)} & 
$\chima\to\nutaubar\staub$ & {\tt sw1br(12)}  \\ \hline 
\rule{0mm}{4mm}
$\chima\to\electron\snebar$ & {\tt sw1br(6)} & 
$\chima\to\bq\stopabar$ & {\tt sw1br(13)}  \\ \hline 
\rule{0mm}{4mm}
$\chima\to\muon\snmubar$ & {\tt sw1br(7)} & 
 &   \\ \hline 
\end{tabular}
\end{center}

\begin{center}
Table 3 \ Decay processes of lighter chargino $\chima$ in \susy23.
\end{center}

\vspace{3mm}

\begin{center}
\begin{tabular}{|l|l||l|l|} \hline
process & abbrev. &
process & abbrev. \\ \hline \hline
\rule{0mm}{4mm}
$\chiob\to\electron\positron\chioa$ & {\tt sn2br(1)} & 
$\chiob\to\antitau\staua$ & {\tt sn2br(16)}  \\ \hline 
\rule{0mm}{4mm}
$\chiob\to\muon\antimuon\chioa$ & {\tt sn2br(2)} & 
$\chiob\to\taum\stauabar$ & {\tt sn2br(17)}  \\ \hline 
\rule{0mm}{4mm}
$\chiob\to\taum\antitau\chioa$ & {\tt sn2br(3)} & 
$\chiob\to\positron\sel$ & {\tt sn2br(18)}  \\ \hline 
\rule{0mm}{4mm}
$\chiob\to\nue\nuebar\chioa$ & {\tt sn2br(4)} & 
$\chiob\to\electron\selbar$ & {\tt sn2br(19)}  \\ \hline 
\rule{0mm}{4mm}
$\chiob\to\numu\numubar\chioa$ & {\tt sn2br(5)} & 
$\chiob\to\antimuon\sml$ & {\tt sn2br(20)}  \\ \hline 
\rule{0mm}{4mm}
$\chiob\to\nutau\nutaubar\chioa$ & {\tt sn2br(6)} & 
$\chiob\to\muon\smlbar$ & {\tt sn2br(21)}  \\ \hline 
\rule{0mm}{4mm}
$\chiob\to\uq\ubar\chioa$ & {\tt sn2br(7)} & 
$\chiob\to\antitau\staub$ & {\tt sn2br(22)}  \\ \hline 
\rule{0mm}{4mm}
$\chiob\to\dq\dbar\chioa$ & {\tt sn2br(8)} & 
$\chiob\to\taum\staubbar$ & {\tt sn2br(23)}  \\ \hline 
\rule{0mm}{4mm}
$\chiob\to\cq\cbar\chioa$ & {\tt sn2br(9)} & 
$\chiob\to\nue\snebar$ & {\tt sn2br(24)}  \\ \hline 
\rule{0mm}{4mm}
$\chiob\to\sq\sbar\chioa$ & {\tt sn2br(10)} & 
$\chiob\to\nuebar\sne$ & {\tt sn2br(25)}  \\ \hline 
\rule{0mm}{4mm}
$\chiob\to\bq\bbar\chioa$ & {\tt sn2br(11)} & 
$\chiob\to\numu\snmubar$ & {\tt sn2br(26)}  \\ \hline 
\rule{0mm}{4mm}
$\chiob\to\positron\ser$ & {\tt sn2br(12)} & 
$\chiob\to\numubar\snmu$ & {\tt sn2br(27)}  \\ \hline 
\rule{0mm}{4mm}
$\chiob\to\electron\serbar$ & {\tt sn2br(13)} & 
$\chiob\to\nutau\sntaubar$ & {\tt sn2br(28)}  \\ \hline 
\rule{0mm}{4mm}
$\chiob\to\antimuon\smr$ & {\tt sn2br(14)} & 
$\chiob\to\nutaubar\sntau$ & {\tt sn2br(29)}  \\ \hline 
\rule{0mm}{4mm}
$\chiob\to\muon\smrbar$ & {\tt sn2br(15)} & 
    &      \\ \hline 
\end{tabular}
\end{center}

\begin{center}
Table 4 \ Decay processes of second lightest neutralino $\chiob$ in \susy23.
\end{center}

\vspace{3mm}

\begin{center}
\begin{tabular}{|l|c|c|} \hline
process  & $\ell = e$ & $\ell = \mu$  \\ \hline \hline
\rule{0mm}{4mm}
$\sll\to\ell\chioa$ & {\tt selbr(1)}& {\tt smlbr(1)}
 \\ \hline 
\rule{0mm}{4mm}
$\sll\to\ell\chiob$ & {\tt selbr(2)}& {\tt smlbr(2)}
 \\ \hline 
\rule{0mm}{4mm}
$\sll\to\nul\chima$ & {\tt selbr(3)}& {\tt smlbr(3)}
 \\ \hline 
\end{tabular}
\end{center}

\begin{center}
Table 5 \ Decay processes of left-handed charged sleptons $\sll$ in 
\susy23.
\end{center}

\vspace{3mm}

\begin{center}
\begin{tabular}{|l|c|c|} \hline
process & $\ell = e$ & $\ell = \mu$ \\ \hline \hline
\rule{0mm}{4mm}
$\slr\to\ell\chioa$ & {\tt serbr(1)}& {\tt smrbr(1)}
 \\ \hline 
\rule{0mm}{4mm}
$\slr\to\ell\chiob$ & {\tt serbr(2)}& {\tt smrbr(2)}
 \\ \hline 
\end{tabular}
\end{center}

\begin{center}
Table 6 \ Decay processes of right-handed charged sleptons $\slr$ 
in \susy23.
\end{center}
\vspace{3mm}

\newpage
\begin{center}
\begin{tabular}{|l|l|} \hline
process  & abbrev.  \\ \hline \hline
\rule{0mm}{4mm}
$\staua\to\tau^-\chioa$ & {\tt sa1br(1)}
 \\ \hline 
\rule{0mm}{4mm}
$\staua\to\tau^-\chiob$ & {\tt sa1br(2)}
 \\ \hline 
\rule{0mm}{4mm}
$\staua\to\nu_{\tau}\chima$ & {\tt sa1br(3)}
 \\ \hline 
\rule{0mm}{4mm}
$\staub\to\tau^-\chioa$ & {\tt sa2br(1)}
 \\ \hline 
\rule{0mm}{4mm}
$\staub\to\tau^-\chiob$ & {\tt sa2br(2)}
 \\ \hline 
\rule{0mm}{4mm}
$\staub\to\nu_{\tau}\chima$ & {\tt sa2br(3)}
 \\ \hline 
\end{tabular}
\end{center}

\begin{center}
Table 7 \ Decay processes of lighter and heavier staus $\stau$ 
in \susy23.
\end{center}

\vspace{3mm}

\begin{center}
\begin{tabular}{|l|c|c|c|} \hline
process & $\ell = e$ & $\ell = \mu$ & $\ell = \tau$ \\ \hline \hline
\rule{0mm}{4mm}
$\snl\to\nul\chioa$ & {\tt snebr(1)}& {\tt snmbr(1)}& {\tt snabr(1)}
 \\ \hline 
\rule{0mm}{4mm}
$\snl\to\nul\chiob$ & {\tt snebr(2)}& {\tt snmbr(2)}& {\tt snabr(2)}
 \\ \hline 
\rule{0mm}{4mm}
$\snl\to\ell\chipa$ & {\tt snebr(3)}& {\tt snmbr(3)}& {\tt snabr(3)}
 \\ \hline 
\end{tabular}
\end{center}

\begin{center}
Table 8 \ Decay processes of sneutrinos $\snl$ in \susy23.
\end{center}

\vspace{3mm}

\begin{center}
\begin{tabular}{|l|l|} \hline
process & abbrev. \\ \hline \hline
\rule{0mm}{4mm}
$\stopa\to\cq\chioa$ & {\tt st1br(1)}
 \\ \hline 
\rule{0mm}{4mm}
$\stopa\to\cq\chiob$ & {\tt st1br(2)}
 \\ \hline 
\rule{0mm}{4mm}
$\stopa\to\bq\chipa$ & {\tt st1br(3)}
 \\ \hline 
\rule{0mm}{4mm}
$\sbottoma\to\bq\chioa$ & {\tt sb1br(1)}
 \\ \hline 
\rule{0mm}{4mm}
$\sbottoma\to\bq\chiob$ & {\tt sb1br(2)}
 \\ \hline 
\end{tabular}
\end{center}

\begin{center}
Table 9 \ Decay processes of 3rd generation squarks $\stopa$ and 
$\sbottoma$ in \susy23.
\end{center}

%% file: appd.tex
\leftline{\large\bf Appendix D. Installation}

The source code  is available by {\tt anonymous ftp} from
{\tt ftp.kek.jp} in the directory \\ {\tt kek/minami/susy23}.
The \susy23 system contains the complete set of Fortran sources
for 23 processes, the three libraries, i.e., 
{\tt BASES/SPRING}, {\tt CHANEL} and
utilities for kinematics.
Those source codes are written in FORTRAN77.
In addition, \susy23 provides the
interface program to generate a few Fortran source files
according to the control data specified by the user.
This program is written in C, YACC and LEX.
\susy23 has been developed on HP-UX, but should run on
any UNIX platform with a fortran complier. 

The procedure of installation is as follows:
\begin{enumerate}
\item Editing {\it Makefile}.

The following macros in {\it Makefile} should be taken care of by 
users themselves.
For example, {\tt SU23DIR} defines the directory name
where \susy23 is installed.
The values of {\tt FC} and {\tt FOPT} define the relevant compiler
name and option for your system.
The other macros can be left as they are.

\begin{tabular}{lcl}
  {\tt SU23DIR} &{\tt =}& directory where \susy23 are installed. \\
  {\tt PRCDIR}   &{\tt =}& directory where process files are installed. \\
 & &  (default is \verb+$(SU23DIR)/prc+.) \\
  {\tt LIBDIR}   &{\tt =}& directory where libraries are installed. \\
 & &  (default is \verb+$(SU23DIR)/lib+.) \\
  {\tt BINDIR}   &{\tt =}& directory where an executable is installed. \\
 & & (default is \verb+$(SU23DIR)/bin+.) \\
  {\tt MACHINE} &{\tt =}& {\tt [hpux|hiux|sgi|dec|sun]} \\
  {\tt FC}      &{\tt =}& FORTRAN compiler command name. \\
  {\tt FOPT}    &{\tt =}& FORTRAN compiler options. \\
  {\tt CERNLIBS} &{\tt =}& CERNLIB including the jetset library. \\
& & For HP-UX, \verb+-L/cern/pro/lib -ljetset74+ \\
& & \verb+-lpacklib -lkernlib -L/lib/pa1.1/ -lm+ \\
\end{tabular}

\item Compilation.

  By executing command {\tt make install} the executable of the interface
  program({\tt susy23}) is generated at {\tt BINDIR}.
  Furthermore six libraries, {\tt BASES/SPRING}, {\tt CHANEL},
  kinematics utility library and three susy libraries,
  are generated in {\tt LIBDIR}.

%
\end{enumerate}

The sample control data files will be found in the directory
{\tt sample}.

%% file: testrun.tex
\leftline{\bf TEST RUN OUTPUT}

\leftline{\tt control data}

\begin{quote}\begin{verbatim}
Process = Eselsz1
Energy  = 200.0d0
type    = tree
itmx    = 7, 15
ncall   = 10000
hadron = yes
mxevnt = 10000
end
\end{verbatim}\end{quote}

Followings are the output files from {\tt BASES} and {\tt SPRING}.
Only one histogram, the momentum distribution of the particle 3, is
shown since the whole output is too lengthy to be included here.

\newpage
\baselineskip 1pt
\lineskip 0pt
\begin{center}
{\scriptsize
\begin{verbatim}
************************************************************************
*----------------------------------------------------------------------*
*----------------------------------------------------------------------*
*------------------------------------------------222---------333-------*
*----------------------------------------------2222222-----3333333-----*
*----------------------------------------------22----22----3----33-----*
*----SSSS SS-UUU--UUUU-----####-##-YYYY---YYYY-------22---------33-----*
*--SSSSSSSS--UUU--UUUU---########--YYYY---YYYY-------22------3333------*
*--SS    SSS--UU----UU---##----###---YY---YY--------22-------3333------*
*--SSSSSSS----UU----UU---#######-----YY---YY------222-----------33-----*
*----SSSSSSS--UU----UU-----#######---YYY-YY------222------------33-----*
*--SS     SS--UU---UUU---##-----##----YYYYY-----22---22----3----33-----*
*--SSSSSSSS----UUUUUUUU--########-----YYYY-----22222222---3333333------*
*--SSSSSSS------UUU-UUU--#######-------YY------22222222-----333--------*
*-------------------------------------YYY------------------------------*
*----------------------------------YYYYYYY-----------------------------*
*----------------------------------YYYYYYY-----------------------------*
*----------------------------------------------------------------------*
*------------------ matrix element generated by GRACE -----------------*
*------------------ version 2.0 (07) --  1997 JUN. 17 -----------------*
************************************************************************
*---------------Copyright -- Minami Tateya Collaboration --------------*
*------------------- E-mail:susy23@minami.kek.jp ----------------------*
************************************************************************
*   Process      : e+e- --> (3)positron (4)selectronl (5)neutralino1
*                                                                      *
*   CM Energy(GeV):     200.00000                                      *
*                                                                      *
*   Mass (Width)                                                       *
*         W-boson:  80.230(  2.080)  Z-boson:  91.190(  2.491)         *
*         u-quark:    .100           d-quark:   .100                   *
*         c-quark:   1.865           s-quark:   .300                   *
*         t-quark: 174.000           b-quark:  5.000                   *
*                                                                      *
*         alpha  :  1/128.070                                          *
*         alpha_s:   .12000                                            *
*                                                                      *
*----------------------------------------------------------------------*
*    SUSY Parameters                                                   *
*----------------------------------------------------------------------*
*       tan(beta)                        =     -2.00                   *
*       SU(2) gaugino mass(Gev)          =     50.00                   *
*       SUSY Higgs parameter(Gev)        =    150.00                   *
*----------------------------------------------------------------------*
*----------------------------------------------------------------------*
*    SFermions Masses(GeV)                                             *
*----------------------------------------------------------------------*
*         s-electron(L):   95.000    s-electron(R):  90.000            *
*                              s-neutrino-electron:  95.000            *
*             s-muon(L):   95.000        s-muon(R):  90.000            *
*                                  s-neutrino-muon:  95.000            *
*              s-tau(1):   90.000         s-tau(2):  95.000            *
*                                   s-neutrino-tau: 100.000            *
*----------------------------------------------------------------------*
*          s-u-quark(L):  200.000     s-u-quark(R): 200.000            *
*          s-d-quark(L):  200.000     s-d-quark(R): 200.000            *
*          s-c-quark(L):  200.000     s-c-quark(R): 200.000            *
*          s-s-quark(L):  200.000     s-s-quark(R): 200.000            *
*          s-t-quark(1):   80.000     s-t-quark(2): 250.000            *
*          s-b-quark(1):   90.000     s-b-quark(2): 210.000            *
*----------------------------------------------------------------------*
*    SFermions Mixing Angles (rad.)                                    *
*----------------------------------------------------------------------*
*                 s-tau:     .000                                      *
*                 s-top:    1.500         s-bottom:    .300            *
*----------------------------------------------------------------------*
\end{verbatim}}
\end{center}

\begin{center}
{\scriptsize
\begin{verbatim}
*----------------------------------------------------------------------*
*       trlinear couplings                                             *
*                 A_tau:    .3000000E+03                               *
*                   A_u:    .5225048E+02                               *
*                   A_d:   -.1732713E+04                               *
*----------------------------------------------------------------------*
*----------------------------------------------------------------------*
*    Neutralino                                                        *
*----------------------------------------------------------------------*
*       neutralino masses(Gev)           =     27.89,     70.53        *
*                                             159.24,    183.34        *
*                                                                      *
*       neutralino mixing matrix (bino bases)                          *
*                   .952     .284     .104     .036                    *
*                  -.207     .864    -.459    -.018                    *
*                  -.122     .280     .611    -.731                    *
*                  -.187     .308     .637     .681                    *
*                                                                      *
*       Neutralino sign factor                                         *
*           1.00    .00 |  1.00    .00 |  1.00    .00 |   .00   1.00   *
*----------------------------------------------------------------------*
*    Chargino                                                          *
*----------------------------------------------------------------------*
*       chargino masses(GeV)             =     69.60,    181.74        *
*                                                                      *
*       chargino mixing matrix                                         *
*                                                .85,      -.53        *
*                                               1.00,       .09        *
*                                                                      *
*       chargino sign factor            =       1.00                   *
*                                                                      *
*----------------------------------------------------------------------*
*   Calculated Total Decay width(GeV)                                  *
*                                                                      *
*  chargino1  : .364E-04  chargino2  : .521E+01                        *
*  neutralino2: .149E-04  neutralino3: .322E+00  neutralino4: .662E+00 *
*  selectronR : .339E+00  selectronL : .392E+00                        *
*  smuonR     : .339E+00  smuonL     : .392E+00                        *
*  stau1      : .325E+00  stau2      : .367E+00                        *
*  sneu_e     : .272E+00  sneu_mu    : .272E+00  sneu_tau   : .351E+00 *
*  stop1      : .240E-03  sbottom1   : .469E-01                        *
*----------------------------------------------------------------------*
*   Experimental Cuts                                                  *
*        Angle Cuts                                                    *
*             Particle 3      -1.000  --->     1.000                   *
*             Particle 4      -1.000  --->     1.000                   *
*             Particle 5      -1.000  --->     1.000                   *
*        Energy CUT                                                    *
*             Particle 3        .001  --->    62.246                   *
*             Particle 4      95.000  --->   120.618                   *
*             Particle 5      27.888  --->    79.382                   *
*        Invariant mass CUT                                            *
*             Inv. Mass 4-5  122.888  --->   199.999                   *
*             Inv. Mass 3-4   95.001  --->   172.112                   *
*             Inv. Mass 3-5   27.889  --->   105.000                   *
*----------------------------------------------------------------------*
*                                                                      *
*   OPTIONS:                                                           *
*        Calculation:         TREE                                     *
*        Width      :         FIX                                      *
************************************************************************
\end{verbatim}}
\end{center}

\begin{center}
{\scriptsize
\begin{verbatim}
************************************************************************
*   Branching Ratios of Sparticles                                     *
*----------------------------------------------------------------------*
*  chargino 1   (sw1)                                                  *
*----------------------------------------------------------------------*
*   (EL ne sz1): .1774   (MU nm sz1): .1774   (TA nt sz1): .2171       *
*   (uq DQ sz1): .2153   (cq SQ sz1): .2129                            *
*   (EL sne)   : .0000   (MU snm)   : .0000   (TA snt)   : .0000       *
*   (ne SEL)   : .0000   (nm SML)   : .0000                            *
*   (nt SA1)   : .0000   (nt SA2)   : .0000                            *
*   (BQ st1)   : .0000                                                 *
*----------------------------------------------------------------------*
*  neutralino 2 (sz2)                                                  *
*----------------------------------------------------------------------*
*   (EL el sz1): .2747   (MU mu sz1): .2747   (TA ta sz1): .3367       *
*   (NE ne sz1): .0200   (NM nm sz1): .0200   (NT nt sz1): .0136       *
*   (UQ uq sz1): .0198   (DQ dq sz1): .0017                            *
*   (CQ cq sz1): .0194   (SQ sq sz1): .0017   (BQ bq sz1): .0177       *
*   (el seR)   : .0000   (mu smR)   : .0000   (ta sa1)   : .0000       *
*   (el seL)   : .0000   (mu smL)   : .0000   (ta sa2)   : .0000       *
*   (ne sne)   : .0000   (nm snm)   : .0000   (na sna)   : .0000       *
*----------------------------------------------------------------------*
*  selectron L  (seL)                                                  *
*----------------------------------------------------------------------*
*   (el sz1)   : .5581   (el sz2)   : .1193   (ne sw1)   : .3226       *
*----------------------------------------------------------------------*
*  selectron R  (seR)                                                  *
*----------------------------------------------------------------------*
*   (el sz1)   : .9914   (el sz2)   : .0086                            *
*----------------------------------------------------------------------*
*  smuon L      (smL)                                                  *
*----------------------------------------------------------------------*
*   (mu sz1)   : .5581   (mu sz2)   : .1193   (nm sw1)   : .3226       *
*----------------------------------------------------------------------*
*  smuon R      (smR)                                                  *
*----------------------------------------------------------------------*
*   (mu sz1)   : .9914   (mu sz2)   : .0086                            *
*----------------------------------------------------------------------*
*  stau 1       (sa1)                                                  *
*----------------------------------------------------------------------*
*   (ta sz1)   : .6223   (ta sz2)   : .1007   (na sw1)   : .2770       *
*----------------------------------------------------------------------*
*  stau 2       (sa2)                                                  *
*----------------------------------------------------------------------*
*   (ta sz1)   : .9883   (ta sz2)   : .0115   (na sw1)   : .0002       *
*----------------------------------------------------------------------*
*  snu_e        (sne)                                                  *
*----------------------------------------------------------------------*
*   (ne sz1)   : .0669   (ne sz2)   : .2900   (el sw1)   : .6431       *
*----------------------------------------------------------------------*
*  snu_mu       (snm)                                                  *
*----------------------------------------------------------------------*
*   (nm sz1)   : .0669   (nm sz2)   : .2900   (mu sw1)   : .6431       *
*----------------------------------------------------------------------*
*  snu_tau      (sna)                                                  *
*----------------------------------------------------------------------*
*   (nt sz1)   : .0555   (nt sz2)   : .2960   (tu sw1)   : .6485       *
*----------------------------------------------------------------------*
*  stop 1       (st1)                                                  *
*----------------------------------------------------------------------*
*   (cq sz1)   : .0000   (cq sz2)   : .0000   (bq sw1)   :1.0000       *
*----------------------------------------------------------------------*
*  sbottom 1    (sb1)                                                  *
*----------------------------------------------------------------------*
*   (bq sz1)   : .1410   (bq sz2)   : .8590                            *
************************************************************************
\end{verbatim}}
\end{center}

\begin{center}
{\scriptsize
\begin{verbatim}
                                                       Date: 97/ 7/ 1  11:49
        **********************************************************
        *                                                        *
        *     BBBBBBB     AAAA     SSSSSS   EEEEEE   SSSSSS      *
        *     BB    BB   AA  AA   SS    SS  EE      SS    SS     *
        *     BB    BB  AA    AA  SS        EE      SS           *
        *     BBBBBBB   AAAAAAAA   SSSSSS   EEEEEE   SSSSSS      *
        *     BB    BB  AA    AA        SS  EE            SS     *
        *     BB    BB  AA    AA  SS    SS  EE      SS    SS     *
        *     BBBB BB   AA    AA   SSSSSS   EEEEEE   SSSSSS      *
        *                                                        *
        *                   BASES Version 5.1                    *
        *           coded by S.Kawabata KEK, March 1994          *
        **********************************************************

     <<   Parameters for BASES    >>

      (1) Dimensions of integration etc.
          # of dimensions :    Ndim    =        5   ( 50 at max.)
          # of Wilds      :    Nwild   =        5   ( 15 at max.)
          # of sample points : Ncall   =     9375(real)    10000(given)
          # of subregions    : Ng      =       50 / variable
          # of regions       : Nregion =        5 / variable
          # of Hypercubes    : Ncube   =     3125

      (2) About the integration variables
          ------+---------------+---------------+-------+-------
              i       XL(i)           XU(i)       IG(i)   Wild
          ------+---------------+---------------+-------+-------
              1     .000000E+00    1.000000E+00     1      yes
              2     .000000E+00    1.000000E+00     1      yes
              3     .000000E+00    1.000000E+00     1      yes
              4     .000000E+00    1.000000E+00     1      yes
              5     .000000E+00    1.000000E+00     1      yes
          ------+---------------+---------------+-------+-------

      (3) Parameters for the grid optimization step
          Max.# of iterations: ITMX1 =        7
          Expected accuracy  : Acc1  =    .1000 %

      (4) Parameters for the integration step
          Max.# of iterations: ITMX2 =       15
          Expected accuracy  : Acc2  =    .0500 %

     <<   Computing Time Information   >>

               (1) For BASES              H: M:  Sec
                   Overhead           :   0: 0: 7.46
                   Grid Optim. Step   :   0:13:41.90
                   Integration Step   :   0:29:13.41
                   Go time for all    :   0:43: 2.77

               (2) Expected event generation time
                   Expected time for 1000 events :     17.81 Sec
\end{verbatim}}
\end{center}

\newpage
\begin{center}
{\scriptsize
\begin{verbatim}
                                                       Date: 97/ 7/ 1  11:49
               Convergency Behavior for the Grid Optimization Step
 ------------------------------------------------------------------------------
 <- Result of  each iteration ->  <-     Cumulative Result     -> < CPU  time >
  IT Eff R_Neg   Estimate  Acc %  Estimate(+- Error )order  Acc % ( H: M: Sec )
 ------------------------------------------------------------------------------
   1 100   .00  1.351E-01  5.295  1.351125(+- .071544)E-01  5.295   0: 1:57.29
   2 100   .00  1.456E-01  1.470  1.447336(+- .020502)E-01  1.417   0: 3:54.83
   3 100   .00  1.430E-01  1.211  1.437373(+- .013232)E-01   .921   0: 5:52.58
   4 100   .00  1.448E-01  1.130  1.441760(+- .010290)E-01   .714   0: 7:50.20
   5 100   .00  1.453E-01  1.205  1.444741(+- .008872)E-01   .614   0: 9:47.20
   6 100   .00  1.457E-01  1.171  1.447447(+- .007871)E-01   .544   0:11:44.74
   7 100   .00  1.452E-01  1.264  1.448097(+- .007233)E-01   .499   0:13:41.90
 ------------------------------------------------------------------------------
\end{verbatim}}
\end{center}

\bigskip
\begin{center}
{\scriptsize
\begin{verbatim}
                                                       Date: 97/ 7/ 1  11:49
               Convergency Behavior for the Integration Step      
 ------------------------------------------------------------------------------
 <- Result of  each iteration ->  <-     Cumulative Result     -> < CPU  time >
  IT Eff R_Neg   Estimate  Acc %  Estimate(+- Error )order  Acc % ( H: M: Sec )
 ------------------------------------------------------------------------------
   1 100   .00  1.467E-01  1.159  1.467078(+- .017002)E-01  1.159   0:15:39.79
   2 100   .00  1.436E-01  1.111  1.450578(+- .011636)E-01   .802   0:17:37.65
   3 100   .00  1.469E-01  1.101  1.456866(+- .009447)E-01   .648   0:19:35.01
   4 100   .00  1.450E-01  1.119  1.455106(+- .008164)E-01   .561   0:21:31.73
   5 100   .00  1.453E-01  1.100  1.454677(+- .007271)E-01   .500   0:23:28.23
   6 100   .00  1.457E-01  1.158  1.454980(+- .006677)E-01   .459   0:25:24.70
   7 100   .00  1.450E-01  1.146  1.454232(+- .006195)E-01   .426   0:27:21.81
   8 100   .00  1.446E-01  1.091  1.453082(+- .005766)E-01   .397   0:29:18.61
   9 100   .00  1.484E-01  1.199  1.455997(+- .005485)E-01   .377   0:31:15.24
  10 100   .00  1.443E-01  1.202  1.454854(+- .005230)E-01   .359   0:33:12.25
  11 100   .00  1.453E-01  1.195  1.454674(+- .005008)E-01   .344   0:35: 8.71
  12 100   .00  1.481E-01  1.212  1.456559(+- .004824)E-01   .331   0:37: 5.18
  13 100   .00  1.457E-01  1.248  1.456616(+- .004662)E-01   .320   0:39: 2.19
  14 100   .00  1.462E-01  1.212  1.456987(+- .004509)E-01   .309   0:40:58.71
  15 100   .00  1.464E-01  1.360  1.457306(+- .004398)E-01   .302   0:42:55.31
 ------------------------------------------------------------------------------
\end{verbatim}}
\end{center}

\newpage
\begin{center}
{\scriptsize
\begin{verbatim}
 Histogram (ID =  1 ) for Momentum of Particle 3                                          
                              Linear Scale indicated by "*"
     x      d(Sigma)/dx      .0E+00      1.5E-03      3.0E-03     4.5E-03      
 +-------+------------------+------------+------------+-----------+-----------+
 I  E  1 I  .000        E  0I                                                 I
 I  .000 I 2.313+- .443 E -5I*OOOOOO                                          I
 I  .150 I 5.546+-1.029 E -5I*OOOOOOOOOOOOO                                   I
 I  .300 I 8.262+-1.265 E -5I*OOOOOOOOOOOOOOOO                                I
 I  .450 I 7.249+-1.200 E -5I*OOOOOOOOOOOOOOO                                 I
 I  .600 I 9.100+-1.624 E -5I*OOOOOOOOOOOOOOOOO                               I
 I  .750 I 8.764+-1.639 E -5I*OOOOOOOOOOOOOOOOO                               I
 I  .900 I 1.090+- .189 E -4I*OOOOOOOOOOOOOOOOOOO                             I
 I 1.050 I 6.979+-1.197 E -5I*OOOOOOOOOOOOOOO                                 I
 I 1.200 I 7.721+-1.723 E -5I*OOOOOOOOOOOOOOOO                                I
 I 1.350 I 9.019+-1.836 E -5I*OOOOOOOOOOOOOOOOO                               I
 I 1.500 I 5.955+-1.129 E -5I*OOOOOOOOOOOOOO                                  I
 I 1.650 I 1.232+- .199 E -4I**OOOOOOOOOOOOOOOOOOO                            I
 I 1.800 I 1.104+- .197 E -4I*OOOOOOOOOOOOOOOOOOO                             I
 I 1.950 I 1.120+- .157 E -4I*OOOOOOOOOOOOOOOOOOO                             I
 I 2.100 I 1.144+- .180 E -4I*OOOOOOOOOOOOOOOOOOO                             I
 I 2.250 I 1.399+- .159 E -4I**OOOOOOOOOOOOOOOOOOOO                           I
 I 2.400 I 1.648+- .104 E -4I**OOOOOOOOOOOOOOOOOOOOO                          I
 I 2.550 I 3.776+- .622 E -4I****OOOOOOOOOOOOOOOOOOOOOOOOOO                   I
 I 2.700 I 2.477+- .081 E -3I*********************OOOOOOOOOOOOOOOOOOOOOOOO    I
 I 2.850 I 2.739+- .064 E -3I***********************OOOOOOOOOOOOOOOOOOOOOOO   I
 I 3.000 I 3.066+- .071 E -3I**************************OOOOOOOOOOOOOOOOOOOOO  I
 I 3.150 I 4.264+- .083 E -3I************************************OOOOOOOOOOOOOI
 I 3.300 I 4.362+- .080 E -3I*************************************OOOOOOOOOOOOI
 I 3.450 I 4.349+- .079 E -3I*************************************OOOOOOOOOOOOI
 I 3.600 I 4.384+- .070 E -3I*************************************OOOOOOOOOOOOI
 I 3.750 I 4.491+- .074 E -3I**************************************OOOOOOOOOOOO
 I 3.900 I 4.467+- .076 E -3I**************************************OOOOOOOOOOOO
 I 4.050 I 4.410+- .071 E -3I*************************************OOOOOOOOOOOOI
 I 4.200 I 4.453+- .069 E -3I**************************************OOOOOOOOOOOO
 I 4.350 I 4.551+- .080 E -3I**************************************OOOOOOOOOOOO
 I 4.500 I 4.469+- .080 E -3I**************************************OOOOOOOOOOOO
 I 4.650 I 4.402+- .077 E -3I*************************************OOOOOOOOOOOOI
 I 4.800 I 4.381+- .082 E -3I*************************************OOOOOOOOOOOOI
 I 4.950 I 4.554+- .084 E -3I**************************************OOOOOOOOOOOO
 I 5.100 I 4.503+- .076 E -3I**************************************OOOOOOOOOOOO
 I 5.250 I 4.564+- .081 E -3I***************************************OOOOOOOOOOO
 I 5.400 I 4.536+- .075 E -3I**************************************OOOOOOOOOOOO
 I 5.550 I 4.380+- .071 E -3I*************************************OOOOOOOOOOOOI
 I 5.700 I 4.602+- .066 E -3I***************************************OOOOOOOOOOO
 I 5.850 I 4.264+- .071 E -3I************************************OOOOOOOOOOOOOI
 I 6.000 I 2.521+- .062 E -3I**********************OOOOOOOOOOOOOOOOOOOOOOO    I
 I 6.150 I 5.346+- .292 E -5I*OOOOOOOOOOOOO                                   I
 I 6.300 I  .000+- .000 E  0I                                                 I
 I 6.450 I  .000+- .000 E  0I                                                 I
 I 6.600 I  .000+- .000 E  0I                                                 I
 I 6.750 I  .000+- .000 E  0I                                                 I
 I 6.900 I  .000+- .000 E  0I                                                 I
 I 7.050 I  .000+- .000 E  0I                                                 I
 I 7.200 I  .000+- .000 E  0I                                                 I
 I 7.350 I  .000+- .000 E  0I                                                 I
 I  E  1 I  .000        E  0I                                                 I
 +-------+------------------+------------------+------------------+-----------+
     x      d(Sigma)/dx     1.0E-05            1.0E-04            1.0E-03      
                              Logarithmic Scale indicated by "O"
\end{verbatim}}
\end{center}

\newpage
\begin{center}
{\scriptsize
\begin{verbatim}

                                                       Date: 97/ 7/ 1  14:28
        **********************************************************
        *                                                        *
        *    SSSSS   PPPPPP   RRRRRR   IIIII  N    NN   GGGGG    *
        *   SS   SS  PP   PP  RR   RR   III   NN   NN  GG   GG   *
        *   SS       PP   PP  RR   RR   III   NNN  NN  GG        *
        *    SSSSS   PPPPPP   RRRRR     III   NNNN NN  GG  GGGG  *
        *        SS  PP       RR  RR    III   NN NNNN  GG   GG   *
        *   SS   SS  PP       RR   RR   III   NN  NNN  GG   GG   *
        *    SSSSS   PP       RR    RR IIIII  NN   NN   GGGGG    *
        *                                                        *
        *                  SPRING Version 5.1                    *
        *           coded by S.Kawabata KEK, March 1994          *
        **********************************************************

     Number of generated events    =     10000
     Generation efficiency         =    23.794 Percent
     Computing time for generation =   443.281 Seconds
                    for Overhead   =     8.240 Seconds
                    for Others     =    10.349 Seconds
     GO time for event generation  =   461.870 Seconds
     Max. number of trials MXTRY   =        50 per event
     Number of miss-generation     =        51 times
\end{verbatim}}
\end{center}

\newpage
\begin{center}
{\scriptsize
\begin{verbatim}
 Original Histogram (ID =  1 ) for Momentum of Particle 3                                      
 Total =     10000 events   "*" : Orig. Dist. in Log Scale.
    x      d(Sig/dx)  dN/dx 1.0E-05            1.0E-04            1.0E-03      
 +-------+----------+-------+------------------+------------------+-----------+
 I  E  1 I  .000E  0I      0I                                                 I
 I  .000 I 2.313E -5I      7I*******    <   O  >                              I
 I  .150 I 5.546E -5I      5I*******<****O* >                                 I
 I  .300 I 8.262E -5I      7I***********<***O* >                              I
 I  .450 I 7.249E -5I     13I**************** <  O >                          I
 I  .600 I 9.100E -5I      3I*<******O***>*****                               I
 I  .750 I 8.764E -5I     10I***************<**O >                            I
 I  .900 I 1.090E -4I      6I*********<****O*>***                             I
 I 1.050 I 6.979E -5I     10I***************<  O >                            I
 I 1.200 I 7.721E -5I     12I*****************< O >                           I
 I 1.350 I 9.019E -5I     10I***************<**O >                            I
 I 1.500 I 5.955E -5I     17I***************     < O >                        I
 I 1.650 I 1.232E -4I      8I************<***O*>**                            I
 I 1.800 I 1.104E -4I      3I*<******O***>*******                             I
 I 1.950 I 1.120E -4I      9I**************<**O*>                             I
 I 2.100 I 1.144E -4I     16I*******************<  O>                         I
 I 2.250 I 1.399E -4I     11I****************<**O*>                           I
 I 2.400 I 1.648E -4I     18I********************<**O>                        I
 I 2.550 I 3.776E -4I     33I**************************<O*>                   I
 I 2.700 I 2.477E -3I    231I******************************************<O>    I
 I 2.850 I 2.739E -3I    262I********************************************<>   I
 I 3.000 I 3.066E -3I    292I********************************************<>*  I
 I 3.150 I 4.264E -3I    416I***********************************************<>I
 I 3.300 I 4.362E -3I    478I*************************************************O
 I 3.450 I 4.349E -3I    472I************************************************<>
 I 3.600 I 4.384E -3I    439I************************************************<>
 I 3.750 I 4.491E -3I    445I************************************************<>
 I 3.900 I 4.467E -3I    457I************************************************<>
 I 4.050 I 4.410E -3I    480I*************************************************O
 I 4.200 I 4.453E -3I    487I*************************************************O
 I 4.350 I 4.551E -3I    451I************************************************<>
 I 4.500 I 4.469E -3I    447I************************************************<>
 I 4.650 I 4.402E -3I    448I************************************************<>
 I 4.800 I 4.381E -3I    469I************************************************<>
 I 4.950 I 4.554E -3I    458I************************************************<>
 I 5.100 I 4.503E -3I    490I*************************************************O
 I 5.250 I 4.564E -3I    471I************************************************<>
 I 5.400 I 4.536E -3I    473I************************************************<>
 I 5.550 I 4.380E -3I    501I*************************************************<
 I 5.700 I 4.602E -3I    488I*************************************************O
 I 5.850 I 4.264E -3I    413I***********************************************<>I
 I 6.000 I 2.521E -3I    223I******************************************<>*    I
 I 6.150 I 5.346E -5I     11I**************  <  O >                           I
 I 6.300 I  .000E  0I      0I                                                 I
 I 6.450 I  .000E  0I      0I                                                 I
 I 6.600 I  .000E  0I      0I                                                 I
 I 6.750 I  .000E  0I      0I                                                 I
 I 6.900 I  .000E  0I      0I                                                 I
 I 7.050 I  .000E  0I      0I                                                 I
 I 7.200 I  .000E  0I      0I                                                 I
 I 7.350 I  .000E  0I      0I                                                 I
 I  E  1 I  .000E  0I      0I                                                 I
 +-------+----------+-------+------------------+------------------+-----------+
    x      d(Sig/dx)  dN/dx    "O" : Generated Events.( Arbitrary unit in Log )
\end{verbatim}}
\end{center}
\pagestyle{empty}